\documentclass[twocolumn]{aastex631}

\usepackage{color, hyperref}
\usepackage{amsmath}
\usepackage{xspace}

\newcommand{\niia}{[N\,{\sc ii}]$_{6548}$\xspace}
\newcommand{\niib}{[N\,{\sc ii}]$_{6585}$\xspace}

\newcommand{\oiiib}{[O\,{\sc iii}]$_{5008}$\xspace}

\newcommand{\ha}{H$\alpha$\xspace}
\newcommand{\hb}{H$\beta$\xspace}
\newcommand{\cii}{[C\,{\sc ii}]$_{158}$\xspace}
\newcommand{\hii}{H\,{\sc ii}\xspace}

\newcommand{\hst}{{\it HST}\xspace}
\newcommand{\jwst}{{\it JWST}\xspace}

\newcommand{\unit}[1]{\mathrm{#1}}

\newcommand{\s}{\unit{s}}

\newcommand{\estar}{E(B-V)_{\mathrm{star}}}
\newcommand{\eneb}{E(B-V)_{\mathrm{neb}}}
\newcommand{\mstar}{M_{\mathrm{star}}}

\newcommand{\pro}{\textsc{Prospector}\xspace}

\shortauthors{Tsujita et al.}
\begin{document}

\title{The ALPINE-CRISTAL-JWST Survey: \\ Stellar and nebular dust attenuation of main-sequence galaxies at $z\sim4-6$}

\author[0000-0002-0498-5041]{Akiyoshi Tsujita}
\affiliation{Institute of Astronomy, Graduate School of Science, The University of Tokyo, 2-21-1 Osawa, Mitaka, Tokyo 181-0015, Japan}

\author[0000-0001-7201-5066]{Seiji Fujimoto}
\affiliation{David A. Dunlap Department of Astronomy and Astrophysics, University of Toronto, 50 St. George Street, Toronto, Ontario, M5S 3H4, Canada}
\affiliation{Dunlap Institute for Astronomy and Astrophysics, 50 St. George Street, Toronto, Ontario, M5S 3H4, Canada}
\author[0000-0002-9382-9832]{Andreas Faisst}
\affiliation{Caltech/IPAC, MS314-6, 1200 E. California Boulevard, Pasadena, CA 91125, USA}

\author[0000-0003-0946-6176]{Médéric Boquien}
\affiliation{Université Côte d'Azur, Observatoire de la Côte d'Azur, CNRS, Laboratoire Lagrange, 06000, Nice, France}

\author[0000-0002-8184-5229]{Juno Li} 
\affiliation{International Centre for Radio Astronomy Research (ICRAR), The University of Western Australia, M468, 35 Stirling Highway, Crawley, WA 6009, Australia}

\author[0000-0002-9400-7312]{Andrea Ferrara}
\affil{Scuola Normale Superiore, Piazza dei Cavalieri 7, 50126 Pisa, Italy}

\author[0000-0003-4569-2285]{Andrew J. Battisti}
\affiliation{International Centre for Radio Astronomy Research (ICRAR), The University of Western Australia, M468, 35 Stirling Highway, Crawley, WA 6009, Australia}
\affiliation{Research School of Astronomy and Astrophysics, Australian National University, Cotter Road, Weston Creek, ACT 2611, Australia}
\affiliation{ARC Center of Excellence for All Sky Astrophysics in 3 Dimensions (ASTRO 3D), Australia}

\author[0009-0007-7842-9930]{Poulomi Dam}
\affiliation{Dipartimento di Fisica e Astronomia Galileo Galilei Universit{\`a} degli Studi di Padova, Vicolo dell’Osservatorio 3, 35122 Padova, Italy}

\author[0000-0002-6290-3198]{Manuel Aravena}
\affiliation{Instituto de Estudios Astrof\'{\i}sicos, Facultad de Ingenier\'{\i}a y Ciencias, Universidad Diego Portales, Av. Ej\'ercito 441, Santiago 8370191, Chile}
\affiliation{Millenium Nucleus for Galaxies (MINGAL), Av. Ej\'ercito 441, Santiago 8370191, Chile}

\author[0000-0002-3915-2015]{Matthieu B\'ethermin}
\affiliation{Universit\'e de Strasbourg, CNRS, Observatoire astronomique de Strasbourg, UMR 7550, 67000 Strasbourg, France}

\author[0000-0002-0930-6466]{Caitlin M. Casey}
\affiliation{Department of Physics, University of California, Santa Barbara, CA 93106, USA}
\affiliation{Cosmic Dawn Center (DAWN), Denmark}

\author[0000-0003-3881-1397]{Olivia R. Cooper}
\affiliation{The University of Texas at Austin, 2515 Speedway Boulevard Stop C1400, Austin, TX 78712, USA}

\author[0000-0001-8519-1130]{Steven L. Finkelstein}
\affiliation{Department of Astronomy, The University of Texas at Austin, Austin, TX, USA}

\author[0000-0002-9122-1700]{Michele Ginolfi}
\affiliation{Universit\`a di Firenze, Dipartimento di Fisica e Astronomia, via G. Sansone 1, 50019 Sesto Fiorentino, Florence, Italy}
\affiliation{INAF -- Arcetri Astrophysical Observatory, Largo E. Fermi 5, I-50125, Florence, Italy}

\author[0009-0008-1835-7557]{Diego A. G\'{o}mez-Espinoza}
\affiliation{Instituto de F\'{i}sica y Astronom\'{i}a, Universidad de Valpara\'{i}so, Avda. Gran Breta\~{n}a 1111, Valpara\'{i}so, Chile}
\affiliation{Millenium Nucleus for Galaxies (MINGAL), Av. Ej\'ercito 441, Santiago 8370191, Chile}

\author[0009-0003-3097-6733]{Ali Hadi}
\affiliation{Department of Physics and Astronomy, University of California, Riverside, 900 University Ave, Riverside, CA 92521, USA}

\author[0000-0002-2775-0595]{Rodrigo Herrera-Camus}
\affiliation{Departamento de Astronomía, Universidad de Concepción, Barrio Universitario, Concepción, Chile}
\affiliation{Millenium Nucleus for Galaxies (MINGAL), Av. Ej\'ercito 441, Santiago 8370191, Chile}

\author[0009-0008-9801-2224]{Edo Ibar}
\affiliation{Instituto de F\'{i}sica y Astronom\'{i}a, Universidad de Valpara\'{i}so, Avda. Gran Breta\~{n}a 1111, Valpara\'{i}so, Chile}
\affiliation{Millenium Nucleus for Galaxies (MINGAL), Av. Ej\'ercito 441, Santiago 8370191, Chile}

\author[0000-0003-4268-0393]{Hanae Inami}
\affiliation{Hiroshima Astrophysical Science Center, Hiroshima University, 1-3-1 Kagamiyama, Higashi-Hiroshima, Hiroshima 739-8526, Japan}

\author[0000-0002-0267-9024]{Gareth C. Jones}
\affiliation{Kavli Institute for Cosmology, University of Cambridge, Madingley Road, Cambridge CB3 0HA, UK}
\affiliation{Cavendish Laboratory, University of Cambridge, 19 JJ Thomson Avenue, Cambridge CB3 0HE, UK}

\author[0000-0002-6610-2048]{Anton M. Koekemoer}
\affiliation{Space Telescope Science Institute, 3700 San Martin Drive, Baltimore, MD 21218, USA}

\author[0000-0002-4052-2394]{Kotaro Kohno}
\affiliation{Institute of Astronomy, Graduate School of Science, The University of Tokyo, 2-21-1 Osawa, Mitaka, Tokyo 181-0015, Japan}
\affiliation{Research Center for the Early Universe, Graduate School of Science, The University of Tokyo, 7-3-1 Hongo, Bunkyo-ku, Tokyo 113-0033, Japan}

\author[0000-0002-1428-7036]{Brian C.\ Lemaux}
\affiliation{Gemini Observatory, NSF NOIRLab, 670 N. A'ohoku Place, Hilo, Hawai'i, 96720, USA}
\affiliation{Department of Physics and Astronomy, University of California, Davis, One Shields Ave., Davis, CA 95616, USA}

\author[0000-0002-9252-114X]{Zhaoxuan Liu}
\affiliation{Kavli Institute for the Physics and Mathematics of the Universe (Kavli IPMU, WPI), UTIAS, Tokyo Institutes for Advanced Study, University of Tokyo, Chiba, 277-8583, Japan}
\affiliation{Department of Astronomy, School of Science, The University of Tokyo, 7-3-1 Hongo, Bunkyo, Tokyo 113-0033, Japan}
\affiliation{Center for Data-Driven Discovery, Kavli IPMU (WPI), UTIAS, The University of Tokyo, Kashiwa, Chiba 277-8583, Japan}
\affiliation{Université Paris-Saclay, Université Paris Cité, CEA, CNRS, AIM, F-91191 Gif-sur-Yvette, France}

\author[0000-0001-9419-6355]{Ilse De Looze}
\affiliation{Sterrenkundig Observatorium, Ghent University, Krijgslaan 281 - S9, B-9000 Gent, Belgium}

\author[0000-0001-7300-9450]{Ikki Mitsuhashi}
\affiliation{Department for Astrophysical \& Planetary Science, University of Colorado, Boulder, CO 80309, USA}

\author[0000-0001-5846-4404]{Bahram Mobasher}
\affiliation{Department of Physics and Astronomy, University of California, Riverside, 900 University Ave, Riverside, CA 92521, USA}

\author[0000-0002-8136-8127]{Juan Molina}
\affiliation{Instituto de F\'{i}sica y Astronom\'{i}a, Universidad de Valpara\'{i}so, Avda. Gran Breta\~{n}a 1111, Valpara\'{i}so, Chile}
\affiliation{Millenium Nucleus for Galaxies (MINGAL), Av. Ej\'ercito 441, Santiago 8370191, Chile}

\author[0000-0001-6652-1069]{Ambra Nanni}
\affiliation{National Centre for Nuclear Research, ul. Pasteura 7, 02-093 Warsaw, Poland}
\affiliation{INAF - Osservatorio astronomico d'Abruzzo, Via Maggini SNC, 64100, Teramo, Italy}

\author[0000-0002-7412-647X]{Francesca Pozzi}
\affiliation{University of Bologna – Department of Physics and Astronomy “Augusto Righi” (DIFA), Via Gobetti 93/2, 40129 Bologna, Italy}
\affiliation{INAF-Osservatorio di Astrofisica e Scienza dello Spazio, Via Gobetti 93/3, 40129, Bologna, Italy}

\author[0000-0001-9687-4973]{Naveen A. Reddy}
\affiliation{Department of Physics \& Astronomy, University of California, Riverside, 900 University Avenue, Riverside, CA 92521, USA}

\author[0000-0003-1682-1148]{Monica Relano}
\affiliation{Dept. Física Te\'{o}rica y del Cosmos, Campus de Fuentenueva, Edificio Mecenas, Universidad de Granada, E-18071, Granada, Spain}
\affiliation{Instituto Universitario Carlos I de Física Te\'{o}rica y Computacional, Universidad de Granada, 18071, Granada, Spain}

\author[0000-0002-9415-2296]{Giulia Rodighiero}
\affiliation{Department of Physics and Astronomy, Università degli Studi di Padova, Vicolo dell’Osservatorio 3, I-35122, Padova, Italy}
\affiliation{INAF—Osservatorio Astronomico di Padova, Vicolo dell’Osservatorio 5, I-35122, Padova, Italy}

\author[0000-0002-9948-3916]{Michael Romano}
\affiliation{Max-Planck-Institut für Radioastronomie, Auf dem Hügel 69, 53121 Bonn, Germany}
\affiliation{INAF - Osservatorio Astronomico di Padova, Vicolo dell’Osservatorio 5, I-35122 Padova, Italy}

\author[0000-0002-1233-9998]{David B. Sanders}
\affiliation{Institute for Astronomy, University of Hawaii, 2680 Woodlawn Drive, Honolulu, HI 96822, USA}

\author[0000-0002-0498-8074]{Prasad Sawant}
\affiliation{National Centre for Nuclear Research, ul. Pasteura 7, 02-093 Warsaw, Poland}

\author[0000-0001-6629-0379]{Manuel Solimano}
\affiliation{Instituto de Estudios Astrof\'isicos, Facultad de Ingenier\'ia y Ciencias, Universidad Diego Portales, Av. Ej\'ercito Libertador 441, Santiago 8370191, Chile}

\author[0000-0002-2906-2200]{Laura Sommovigo}
\affiliation{Center for Computational Astrophysics, Flatiron Institute, 162 5th Avenue, New York, NY 10010, USA}

\author[0000-0003-3256-5615]{Justin Spilker}
\affiliation{Department of Physics and Astronomy and George P. and Cynthia Woods Mitchell Institute for Fundamental Physics and Astronomy, Texas A\&M University, 4242}

\author[0000-0001-9728-8909]{Ken-ichi Tadaki}
\affiliation{Faculty of Engineering, Hokkai-Gakuen University, Toyohira-ku, Sapporo 062-8605, Japan}

\author[0000-0002-3258-3672]{Livia Vallini}
\affiliation{INAF – Osservatorio di Astrofisica e Scienza dello Spazio di Bologna, Via Gobetti 93/3, 40129 Bologna, Italy}

\author[0000-0002-5877-379X]{Vicente Villanueva}
\affiliation{Departamento de Astronom{\'i}a, Universidad de Concepci{\'o}n, Barrio Universitario, Concepci{\'o}n, Chile}

\author[0000-0002-7964-6749]{Wuji Wang}
\affiliation{Caltech/IPAC, 1200 E. California Blvd. Pasadena, CA 91125, USA}

\author[0000-0002-2318-301X]{Giovanni Zamorani}
\affiliation{INAF – Osservatorio di Astrofisica e Scienza dello Spazio di Bologna, Via Gobetti 93/3, 40129 Bologna, Italy}

\author{ALPINE+CRISTAL collaborations}

\begin{abstract}
Characterizing dust attenuation is crucial for revealing the intrinsic physical properties of galaxies.
We present an analysis of dust attenuation in 18 spectroscopically confirmed star-forming main-sequence galaxies at $z = 4.4–5.7$ observed with \jwst/NIRSpec IFU and NIRCam, selected from the ALPINE and CRISTAL ALMA large programs. We fit the emission line fluxes from NIRSpec and the broad-band photometry from NIRCam with \pro, using both spatially integrated emission and $\sim0.6$ kpc pixel-by-pixel measurements. 
We derive the stellar-to-nebular dust attenuation ratio ($f=\estar/\eneb$) from the SED fits and the Balmer decrement with \ha and \hb. 
Although individual galaxies show large scatter, the best-fit value is $f = 0.51^{+0.04}_{-0.03}$, slightly higher than that measured for local starburst galaxies.
We find weak correlations of $f$ with galaxy properties, increasing with higher specific star-formation rates, younger stellar ages, and more recent star-formation.
For the range of $\estar = 0.009-0.15$ mag for in our sample, assuming $f = 1$ (often adopted in high-redshift studies) instead of $f = 0.51$ underestimate line luminosities and ionizing photon production efficiency $\xi_\mathrm{ion}$ by $\sim3-36\%$ and $\sim4-46\%$, respectively.
Finally, total stellar masses estimated from spatially-integrated SED fits with a delayed-$\tau$ star-formation histories are systematically smaller than the sums of pixel-by-pixel SED fits by a median of $\sim 0.26$ dex, likely because the integrated fits are biased toward luminous young stellar populations.
\end{abstract}


\section{Introduction} \label{sec:intro}
Dust affects the spectral energy distribution (SED) and emission line intensities of galaxies by absorbing and scattering ultraviolet (UV) and optical emission, primarily produced by young stars and re-emitting the energy at the far-infrared (FIR) and sub-millimeter (sub-mm) wavelengths. 
Accurate and precise measurement and correction of dust attenuation\footnote{In this paper, we use the term attenuation to describe the effective impact of dust on the integrated light from a galaxy, including both dust properties and the geometry of stars and dust. This differs from extinction, which refers to dust absorption and scattering along a single line of sight and is determined solely by dust grain properties \citep[e.g.,][]{Salim2020ARA&A..58..529S}.} are essential for deriving a true picture of galaxy properties, such as stellar mass and star formation rate (SFR) \citep[e.g.,][]{Salim2020ARA&A..58..529S}.
Such corrections to dust attenuation in individual galaxies ultimately impact the global picture on cosmic scales, including key measurements such as the cosmic star formation rate density (CSFRD; \citealt{Madau2014}) and the sources and processes of cosmic reionization. 

The dust distribution in galaxies is often described using a two-component model \citep[e.g.,][]{cf2000ApJ...539..718C}, consisting of dust in the diffuse interstellar medium (ISM) and in birth clouds associated with star-forming regions (e.g., \citealt{Sommovigo2020MNRAS.497..956S}). The birth clouds have a finite lifetime of $\sim 10$--30 Myr (e.g., \citealt{Blitz1980ApJ...238..148B, Chevance2020SSRv..216...50C}). Emission lines from \hii regions and stellar continuum from young stars are produced on shorter timescales ($\sim6$ Myr), and thus are attenuated by both components. In contrast, the stellar continuum from stars that live longer than the birth clouds is only attenuated by the diffuse dust. This two-component model has successfully explained the higher attenuation observed in nebular lines, as inferred from the \ha/\hb ratio, compared to that of the stellar continuum, derived from the UV/optical spectral slope in local galaxies \citep[e.g.,][]{Fanelli1988ApJ...334..665F, Calzetti1994ApJ...429..582C, Calzetti2000, battisti2016characterizing}. \cite{Calzetti2000} quantified this difference as $f=\estar/\eneb=0.44\pm 0.03$ for local starburst galaxies\footnote{This value is derived using two different dust attenuation curves: \cite{Calzetti2000} curve for stellar attenuation and Milky Way extinction curve (\citealt{Fitzpatrick1999PASP..111...63F}) for nebular attenuation. If the \cite{Calzetti2000} curve is used to measure both attenuation, this value becomes $f=0.58$ \citep{Steidel2014ApJ...795..165S}.}, and similarly, \citet{Kreckel2013ApJ...771...62K} found that in star-forming regions within local galaxies, the relationship follows $f=0.470 \pm 0.006$. However, the validity of these relationships at high redshift remains an open question.
Several works argue that the $f$ factor depends on the star formation activity of galaxies, such as SFR and specific SFR (sSFR) as it reflects the fraction of diffuse dust components relative to short-lived dense birth clouds \citep[e.g.,][]{Kashino2013ApJ...777L...8K, Price2014ApJ...788...86P, Reddy2015ApJ...806..259R, Reddy2020ApJ...902..123R, Koyama2019PASJ...71....8K}. 
Additionally, the clumpiness of galaxies, which influences the spatial distribution of stars and dust, may also affect the $f$ factor.
Since these properties exhibit redshift dependence \citep[e.g.,][]{Faisst2016ApJ...821..122F, sattari2023fraction}, it is reasonable to expect the $f$ factor to evolve over cosmic time.
Recent studies \citep[e.g.,][]{Reddy2020ApJ...902..123R, Lorenz2023ApJ...951...29L} further suggest that the relation between stellar and nebular attenuation at high redshift may differ from that observed in local galaxies: while in local galaxies the UV continuum often includes a substantial contribution from older stellar populations affected mainly by diffuse dust, in young high-$z$ galaxies with high sSFRs the UV is still dominated by OB associations. In this case, the difference between nebular and stellar reddening may arise primarily from variations in dust column densities along different sightlines to OB associations, with nebular emission tracing the youngest and dustier regions and the UV continuum arising from slightly older and less obscured ones.

In studies of high-redshift galaxies, it is often the case that only one of either \ha or \hb is available, making it difficult to measure the Balmer decrement.
In such cases, nebular attenuation is often estimated by dividing the stellar reddening derived from the UV slope or SED fitting by the $f$ factor \citep[e.g.,][]{Matthee2017MNRAS.465.3637M}. Thus, the intrinsic nebular line luminosity depends on the assumed $f$ factor, which in turn affects key derived quantities such as the line luminosity function, line-based SFR, and ionizing photon production efficiency $\xi_{\rm ion}$ \citep[e.g.,][]{Faisst2019ApJ...884..133F, Saito2020MNRAS.494..199S}. Accurately constraining the $f$ factor using a representative sample of high-redshift galaxies is therefore critical for ensuring robust derivations and reliable interpretations of these astrophysical properties.

At $z\gtrsim0.5$, the \ha line shifts into less accessible near-infrared window, making Balmer decrement measurements more challenging.
However, even prior to the advent of the {\it James Webb Space Telescope} (\jwst), it was possible to measure the Balmer decrement up to $z \lesssim 2.5$, leading to various conclusions.
Some studies suggest that the differential reddening ratio $f$ is close to unity in $z\sim2$ galaxies \citep[e.g.,][]{Erb2006ApJ...647..128E, Shivaei2015ApJ...804..149S, Pannella2015ApJ...807..141P, Puglisi2016A&A...586A..83P}, while others support consistency with local starbursts as found in the \cite{Calzetti2000} work \citep[e.g.,][]{FS2009ApJ...706.1364F, Yoshikawa2010ApJ...718..112Y, Wuyts2011ApJ...738..106W, Mancini2011ApJ...743...86M, Price2014ApJ...788...86P, Rodriguez2022MNRAS.510.2061R}. Numerous works report intermediate values between the aforementioned measurements \citep[e.g.,][]{Kashino2013ApJ...777L...8K, Wuyts2013ApJ...779..135W, Valentino2015ApJ...801..132V, Pannella2015ApJ...807..141P, Theios2019ApJ...871..128T}. For $z \gtrsim 2.5 $, however, such Balmer decrement observations remained entirely unexplored before \jwst, leaving the $f$ factor unconstrained in this redshift regime.

The advent of \jwst/NIRSpec has revolutionized these measurements by enabling the detection of \ha and \hb up to $z \sim 7$ and $z \sim 9.5$, respectively. This breakthrough allows for direct measurements of the $f$ factor at redshifts previously inaccessible. In this study, we address these challenges by investigating the $f$ factors of star-forming main-sequence (MS) galaxies at $4.4 < z < 5.7$ using \jwst/NIRCam and NIRSpec data. 

This paper is organized as follows. Section \ref{sec:data} describes the observational data and its reduction. In Section \ref{sec:analysis}, we present the SED modeling methods and Balmer decrement measurements. Section \ref{sec:result} presents the obtained $f$ factors, explores the correlation between $f$ and galaxy properties, and discusses the potential impact on key emission-line-derived quantities, such as the line luminosity function, \ha-based SFRD, and $\xi_\mathrm{ion}$. Finally, a summary is provided in Section \ref{sec:summary}. Throughout the paper, we adopt a cosmology with $\Omega_{\mathrm{m}}=0.3$, $\Omega_\Lambda=0.7$, and $H_0=70\,\mathrm{km\,s^{-1}\,Mpc^{-1}}$ and the Chabrier initial mass function (IMF, \citealt{Chabrier2003}). 
All magnitudes in this paper are expressed in the AB system \citep{Oke1974}. 

\section{Observations and Data reduction} \label{sec:data}
\subsection{\jwst/NIRSpec IFU}
We analyze the \jwst/NIRSpec IFU observation data from the Cycle-2 GO program (GO-3045, PI: A. Faisst) -- the {\it ALPINE-CRISTAL JWST survey}. This program observed a total of 18 spectroscopically confirmed star-forming MS galaxies at $z=4.4-5.7$ with \jwst/NIRSpec IFU during April--May 2024. The sample selection was based on galaxies previously studied in ALMA large programs, including ALPINE \citep[e.g.,][]{lefevre2020A&A...643A...1L, bethermin2020A&A...643A...2B, faisst2022Univ....8..314F} and CRISTAL \citep[e.g.,][]{Li2024ApJ...976...70L, Mitsuhashi2024A&A...690A.197M, Ikeda2025A&A...693A.237I, Herrera-Camus2025A&A...699A..80H}, within the COSMOS field. 
Our sample spans stellar masses of \( \log(M_\star/M_\odot) \sim 9.0 \)--11.0 and SFRs of \( \sim10 \)--200 \( M_\odot\,\mathrm{yr}^{-1} \), and is biased toward the massive end of the ALPINE parent population due to selection requirements for high signal-to-noise ratio (SNR) spectroscopy. It is therefore complementary to other JWST surveys that tend to focus on lower-mass galaxies \citep[e.g.,][]{Morishita2024ApJ...971...43M, Nakajima2023ApJS..269...33N, Curti2023MNRAS.518..425C, Sarker2025ApJ...978..136S}. Nevertheless, the sample still lies about two orders of magnitude below the knee of the stellar mass function at \( z \sim 5 \) \citep{Davidzon2017A&A...605A..70D}. For details of the sample selection, we refer the reader to \cite{Faisst2025survey}.

All targets were observed using the NIRSpec disperser-filter combinations G235M/F170LP ($1.7~\mu m \lesssim \lambda \lesssim 3.2~\mu m,  R \sim 1000$) and G395M/F290LP ($2.9~\mu m \lesssim \lambda \lesssim 5.3~\mu m,  R \sim 1000$),  with the exception of DC-842313, which was exclusively observed with G235M/F170LP. Deeper G395H/F290LP observations for DC-842313 were carried out in a separate program (GO-4265, PIs J.~González-López \& M.~Aravena; see also \citealt{Solimano2025A&A...693A..70S}) during the same cycle, and are incorporated into our analysis.
For the GO-3045 program, a 2-point sparse cycling dither pattern was used, with each exposure consisting of $\sim$30--60 groups per integration and 1--3 integrations, leading to total on-source exposure times between
$\sim$500 and 7000~seconds per target. The GO-4265 program employed a 9-point small cycling dither pattern, using $\sim$18 groups per integration and 9 integrations, resulting in an on-source exposure time of 11,948 seconds.
For more information on the {\it ALPINE-CRISTAL IFU Survey}, we refer the reader to \cite{Faisst2025survey}.

We reduce the data using the standard \jwst\ pipeline (version 1.16.0) with the Calibration References Data System (CRDS) context jwst1298.pmap, generally following the procedure of \cite{Rigby2025ApJ...978..108R} with some additional modifications. 
After Stage 1, we apply further corrections to reduce vertical pattern noise and snowball artifacts. Before Stage 3, we inspect all exposures and manually mask leakage from intermittently open MSA shutters. After the pipeline processing, we further refine the data by modeling and subtracting stripe patterns in continuum maps, rescaling error estimates to account for underestimated uncertainties, and correcting astrometry by aligning the IFU data to NIRCam imaging.
Full descriptions of the IFU data reduction is given in \cite{Fujimoto2025arXiv251016116F}.

\subsection{\jwst/NIRCam Imaging}
Most of the IFU targets (17/18) have \jwst/NIRCam imaging data from the COSMOS-Web program (PID: 1727, PI: J. Kartaltepe; e.g., \citealt{Casey2023ApJ...954...31C}), observed with F115W, F150W, F277W, and F444W filters. 
Among the 17 targets with NIRCam data, 4 (DC-630594, DC-742174, VC-5100994794, and VC-5101244930) were also observed through the PRIMER program (PID: 1837, PI: J. Dunlop; e.g., \citealt{Donnan2024MNRAS.533.3222D}), providing additional imaging in the F090W, F200W, F356W, and F410M filters. The combined images are drizzled onto a common grid with a pixel size of 0\arcsec.03. Full descriptions of the NIRCam data reduction are given in Franco et al. (in prep). For additional details on the NIRCam data for our sample, see also \cite{Faisst2025survey}.

For the remaining target (DC-417567), which lacks NIRCam coverage, we generated pseudo-NIRCam images by convolving the NIRSpec IFU data with the NIRCam filter transmission curves. The G235M/F170LP and G395M/F290LP data were used to create pseudo-NIRCam F277W and F444W images, respectively. An analysis of our sample shows that these pseudo-images are consistent with actual NIRCam observations within $\lesssim15\%$ \citep{Fujimoto2025arXiv251016116F}.
In addition to these pseudo-images, this target is also observed with \hst ACS and WFC3, including the F160W filter, as part of the CANDELS survey \citep{Grogin2011ApJS..197...35G, Koekemoer2011ApJS..197...36K}.
For this work, we used the F814W and F160W images, with the reduction and PSF models provided by \cite{Li2024ApJ...976...70L}.


\subsection{ALMA}
Our entire sample was observed as part of the ALPINE ALMA Large Program during Cycle 5 (ID: 2017.1.00428.L; PI: O. Le F\`evre; \citep{lefevre2020A&A...643A...1L}), targeting the \cii line and the underlying continuum at an angular resolution of $\sim 1\arcsec$.
Most of these galaxies (17/18) were subsequently followed up in the CRISTAL ALMA Large Program during Cycle 8 (ID: 2021.1.00280.L; PI: R. Herrera-Camus; \citealt{Herrera-Camus2025A&A...699A..80H}) with higher angular resolution, again targeting the \cii line. The remaining galaxy (DC-873756 or CRISTAL-24) was also observed at a comparable high resolution in a separate ALMA project \citep{Devereaux2024A&A...686A.156D}.

The data were calibrated and flagged in the standard manner using the Common Astronomy Software Applications package ({\tt CASA}; \citealt{CASA2022}).
We subsequently combined visibility data from both compact and extended array configurations and performed imaging using {\tt CASA/tclean} with natural weighting. Since this study focuses solely on the dust continuum, we selected only channels free from \cii emission to create the dust continuum image. The final beam sizes range from $\sim 0.08\arcsec$ to $0.45\arcsec$, with a median resolution of $0.26\arcsec$.
Details of the survey design, data reduction, and data products of CRISTAL are presented in \cite{Herrera-Camus2025A&A...699A..80H}.

Note that ALMA photometry is not included in the SED fitting presented in this paper.
The potential impact of including ALMA photometry is briefly discussed in Appendix~\ref{sec:appe_sed_ALMA}.

\subsection{PSF matching and pixel alignment}\label{sec:PSF_match}
Before performing SED fitting, we match the point spread functions (PSFs) of all observations to that of the F444W filter ($\sim0.16\arcsec$) to eliminate potential artificial color gradients introduced by differences in spatial resolution.
We model the \jwst PSF for both NIRCam and NIRSpec data using the WebbPSF package \citep{webbpsf2014SPIE.9143E..3XP} and apply PSF matching using the kernel derived from the {\tt photutils/create\_matching\_kernel} astropy function.
For kernel construction, we use the {\tt SplitCosineBellWindow} window function with parameters $\alpha = 0.4$ and $\beta = 0.3$ to smoothly taper the kernel edges and suppress ringing artifacts.
All observations are then reprojected to match the World Coordinate System (WCS) and pixel grid of the NIRSpec IFU/G395M data cube, which has the largest pixel scale ($0.1\arcsec$), using the {\tt reproject/reproject\_exact} astropy function.

\subsection{Noise map/cube}
For NIRCam images, we apply {\tt Photutils/Background2D} astropy function to the processed images (section~\ref{sec:PSF_match}) to obtain a map of the RMS background variation.
For NIRSpec IFU data, we first rescale the error cube provided by the pipeline to match the observed RMS levels for each wavelength channel (see \citealt{Fujimoto2025arXiv251016116F} for details). We then apply the same processing steps as in section~\ref{sec:PSF_match} to generate the final error cube.
When extracting photometry and line fluxes, the errors are computed as the square root of the sum of the corresponding noise map/cube.

\section{Analysis}\label{sec:analysis}

\subsection{Aperture for photometry and spectroscopy}\label{sec:mask}
For the pixel-by-pixel SED fitting described in the following section, we define a common pixel mask for both the NIRCam imaging and NIRSpec IFU data by selecting pixels with detections greater than 3$\sigma$ in at least two NIRCam filters (see top left panel of Figure \ref{fig:sed}). The spatially integrated photometry and line fluxes are then calculated as the sum over these selected pixels.
If the identified pixel groups are clearly separated and their associated objects have similar redshifts based on IFU data, we treat them as a galaxy pair and perform integrated photometry separately for each component. Such galaxy pairs are distinguished by adding suffixes “a” or “b” to their source IDs, as listed in Table~\ref{tbl:sed_results}.

To account for diffuse extended emission that may not be captured by the selected pixels, we use the {\tt Photutils/detect\_sources} function on the NIRCam F444W image to create Kron apertures with a Kron factor of 2.5 and a minimum radius of 3 pixels. The photometric measurements within these F444W-based Kron apertures are used to correct for additional flux from extended emission. Furthermore, we calculate the fraction of the PSF enclosed within the Kron aperture and apply a correction for the flux outside the aperture to obtain a more complete total flux estimate.

\subsection{Spatially integrated/resolved line flux measurement}
The NIRSpec IFU data cover major optical emission lines, including [O\,{\sc ii}]$_{3727,3730}$, \hb, [O\,{\sc iii}]$_{4960, 5008}$, \ha, [N\,{\sc ii}]$_{6548, 6585}$, and [S\,{\sc ii}]$_{6716, 6731}$ for all galaxies in our sample, spanning a redshift range of $z = 4.4$--$5.7$. 
To construct spatially resolved intensity maps, we first focus on the H$\alpha$ line, which is partially blended with the [N\,{\sc ii}] doublet.
While per-pixel Gaussian fitting is often employed to deblend emission lines, it becomes unreliable in low SNR regions and may bias the total flux.
To address this, we perform a Gaussian fit to the spatially integrated spectrum within the aperture defined in Section~\ref{sec:mask}, modeling \ha and [N\,{\sc ii}]$_{6548,6585}$ with three Gaussians plus an underlying continuum. The \niia flux is fixed to $1/2.96$ of \niib flux \citep{Galavis1997A&AS..123..159G}, and all lines are constrained to share the same redshift and full width at half maximum (FWHM; in velocity units).
We define an emission line as detected when at least three spectral channels within $\pm5$ channels of the line center have SNR $>3$.
Only lines meeting this criterion are fitted.
We then generate a temporary \ha map by subtracting the continuum (median of nearby line-free channels) and summing over three spectral channels ($\sim$350 km/s) centered on \ha peak channel, where [N\,{\sc ii}] contamination is negligible. This map is rescaled using the ratio of the total H$\alpha$ flux of the best-fit Gaussian to the flux summed over the same three channels of the Gaussian model, yielding the final H$\alpha$ intensity map. We confirm that H$\alpha$ is robustly detected in all galaxies (i.e., satisfying the above criterion). Note that this method assumes uniform line profiles across spatial pixels.

The same approach is applied to other emission lines. For [O\,{\sc iii}]$_{4960,5008}$, we fit both lines simultaneously, fixing their flux ratio to 1:2.98 \citep{OIII2000MNRAS.312..813S}, with redshift and FWHM fixed to the H$\alpha$ values. For emission lines not detected in the integrated spectrum, we instead sum the continuum-subtracted flux over a wavelength window equal to the FWHM derived from the H$\alpha$ Gaussian fit.
Spatially integrated line fluxes are finally computed by summing the pixel values within the photometric aperture.

\begin{figure*}[htb]
\centering
\includegraphics[width=0.9\textwidth]{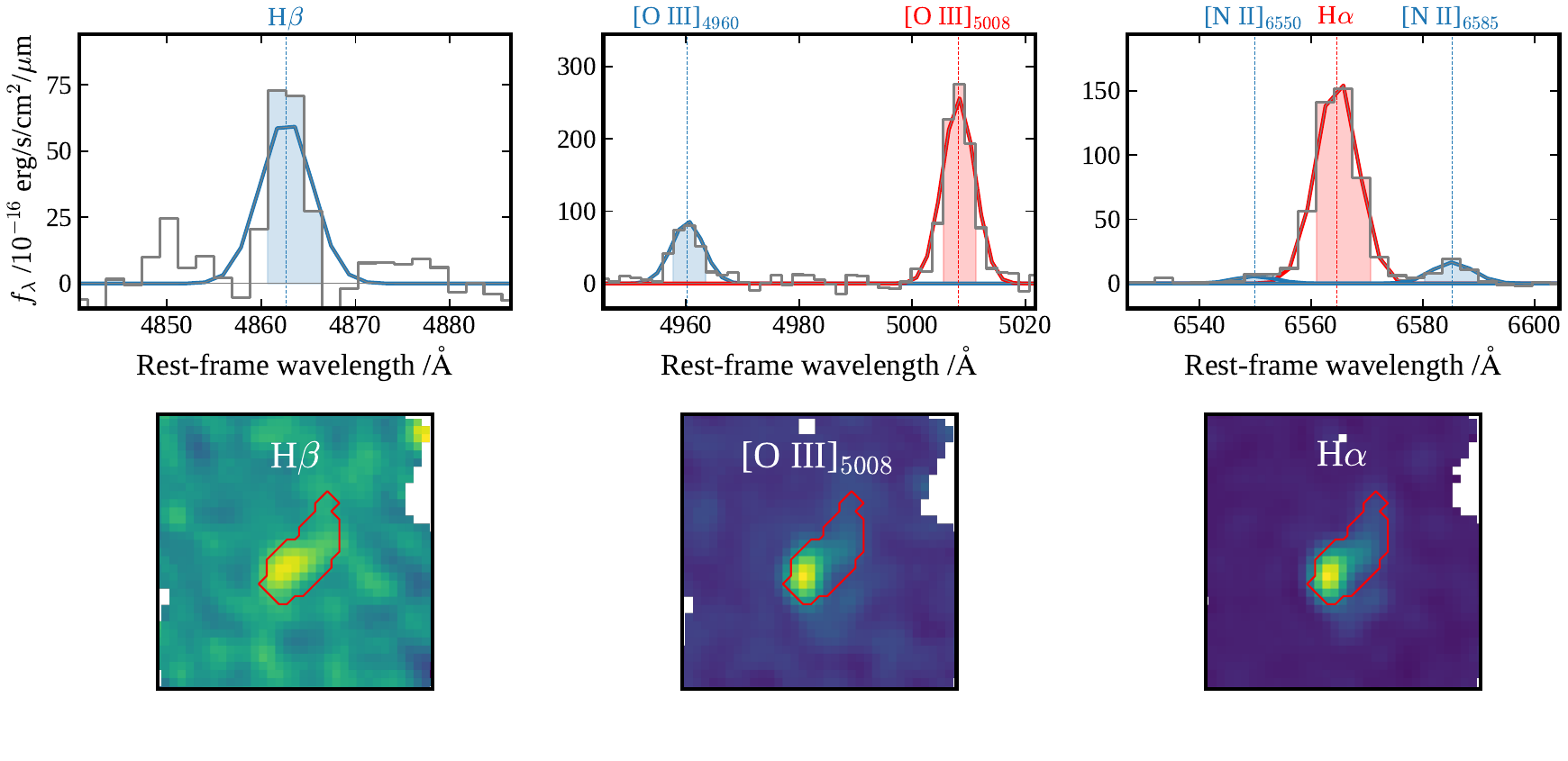}
\caption{
An example of the spatially-integrated spectrum and velocity-integrated emission-line maps for VC-5110377875 at $z=4.55$. (Top) Spectra around H$\beta$, [O\,{\sc iii}]$_{4960,5008}$, and H$\alpha$+[N\,{\sc ii}], with best-fit Gaussian models overlaid in red and blue.
Shaded regions indicate the three-channel windows centered on the line peaks, which are used to construct temporary emission maps.  
(Bottom) Emission maps obtained by summing the continuum-subtracted cube over those shaded channels and rescaled to match the total line flux from the Gaussian fit. The red line is the photometric aperture.
}
\label{fig:specfit}
\end{figure*}

Figure~\ref{fig:specfit} shows an example of the spatially-integrated spectrum fitting and the resulting velocity-integrated emission map.

\subsection{Balmer decrement}
The nebular reddening is calculated from the Balmer decrement assuming the average Milky Way (MW) extinction curve \citep{Cardelli1989ApJ...345..245C}: $\eneb = \frac{2.5}{k_{\mathrm{H}\beta}-k_{\mathrm{H}\alpha}}\log(\frac{\mathrm{H}\alpha/\mathrm{H}\beta}{R_0})$, where $k_{\mathrm{H}\alpha}$ and $k_{\mathrm{H}\beta}$ are the reddening curve values at the wavelengths of H$\alpha$ and H$\beta$, respectively. The constant $R_0$ denotes the intrinsic (unreddened) Balmer decrement. We adopt the standard Case~B recombination, which at an electron temperature of $T_e=10^4$ K and an electron density of $n_e=10^2$ cm$^{-3}$ gives $R_0 = 2.86$ \citep{agn22006agna.book.....O}. 
A more detailed assessment of the $T_e$ and $n_e$ for the ALPINE sample is presented in \cite{Faisst2025survey}.
We note that the derived nebular reddening $\eneb$ does not significantly vary under different assumptions for the attenuation/extinction curve shape, as the shapes of various curves, such as those for the Small Magellanic Cloud (SMC), Large Magellanic Cloud (LMC; \citealt{Gordon2003ApJ...594..279G}), and Calzetti, are very similar in the wavelength range of \ha\ and \hb.
For VC-5100541407, the derived $A_{V,\mathrm{neb}}$ is negative due to an observed H$\alpha$/H$\beta$ ratio below the theoretical value (see Table~\ref{tbl:sed_results}). However, the result is consistent with $A_{V,\mathrm{neb}}=0$ within $1\sigma$ uncertainty.

\subsection{SED fitting}\label{sec:sedfit}
We use the SED-fitting code \pro \citep{Leja2017ApJ...837..170L, Johnson2021ApJS..254...22J} to simultaneously fit the photometric and spectroscopic data of NIRCam and NIRSpec. We utilize the Flexible Stellar Population Synthesis (FSPS) stellar population synthesis models \citep{Conroy2009ApJ...699..486C,Conroy2010ApJ...712..833C,python-fsps2024zndo..12447779J} with the MILES spectral library \citep{MILES2006MNRAS.371..703S}, and MIST isochrones \citep{Dotter2016ApJS..222....8D, Choi2016ApJ...823..102C}. In our spectroscopic data, the continuum was only faintly detected, making full spectrum fitting impractical. 
Instead, we use the {\tt LineSpecModel} class in \pro, incorporating the fluxes of bright emission lines (\ha, \hb, \oiiib, and \niib). 
The line fluxes are incorporated as independent constraints in the fit, in the same way as the photometric fluxes, i.e., they enter the likelihood function as additional “photometric points” with their associated uncertainties.
Including observed line fluxes alongside the photometric data is essential for disentangling the contributions from the stellar continuum and line emission, as well as for providing stronger constraints on the SFH.
To account for calibration uncertainties between NIRCam and NIRSpec fluxes, the {\tt linespec\_scaling} parameter is treated as a free parameter with a Gaussian prior centered at 1 with a standard deviation of 10\%.

Throughout the SED fitting, the redshift is fixed to the spectroscopic redshift determined from the observed \ha line. We adopt both parametric and non-parametric SFH models to fit the spatially integrated emission. The parametric SFH follows a delayed-$\tau$ model ($\mathrm{SFR}(t)\propto t \exp(-t/\tau)$), where the stellar age ($t_{\text{age}}$) ranges from 1 Myr to the cosmic age at the source redshift, and the timescale of the star-formation ($\tau$) varies between 100 Myr and 1000 Gyr (nearly constant SFH). The non-parametric SFH adopts a continuity prior with six time bins. Following \cite{Tacchella2022ApJ...927..170T}, the first bin is fixed at $0-10$ Myr to capture variation in the recent SFH of galaxies, while the other bins are spaced equally in logarithmic time between 10 Myr and a lookback time that corresponds to $z = 20$.
For pixel-by-pixel (spatially resolved) SED fitting, only the parametric delayed-$\tau$ SFH model is used.
This is primarily because the SNR of emission in individual pixels is significantly lower than in the integrated case, making it difficult to robustly constrain non-parametric SFHs. Furthermore, each pixel spans $\sim$0.1 arcsec (corresponding to $\sim$0.6 kpc at $z\sim5$), which is relatively small compared to the galaxy scale. Within such localized regions, it is plausible that the underlying stellar populations share a broadly similar SFH. 

We adopt a single-component dust attenuation model based on the \cite{Calzetti2000} curve, applying a power-law modification to the slope ($\delta$) as a free parameter (flat prior between $-1<\delta<0.4$), while excluding the UV bump.
Although a two-component dust attenuation model is more realistic, deriving $\estar$ from the fit is more challenging since the $\estar$ is a non-trivial combination of the attenuations experienced by both young and old stars. As discussed in detail by \cite{Boquien2022A&A...663A..50B} (see Figure 10), the relative fractions and attenuations of these two populations affect the shape of the attenuation curve.
Allowing $\delta$ to vary mimics the effect of different relative contributions from young and old stellar populations in two-component models.
Therefore, we employ a single-component dust model with $\delta$ free to derive $\estar$ or $A_{V,\text{star}}$.

\begin{figure*}[!htbp]
\centering
\includegraphics[width=0.9\linewidth]{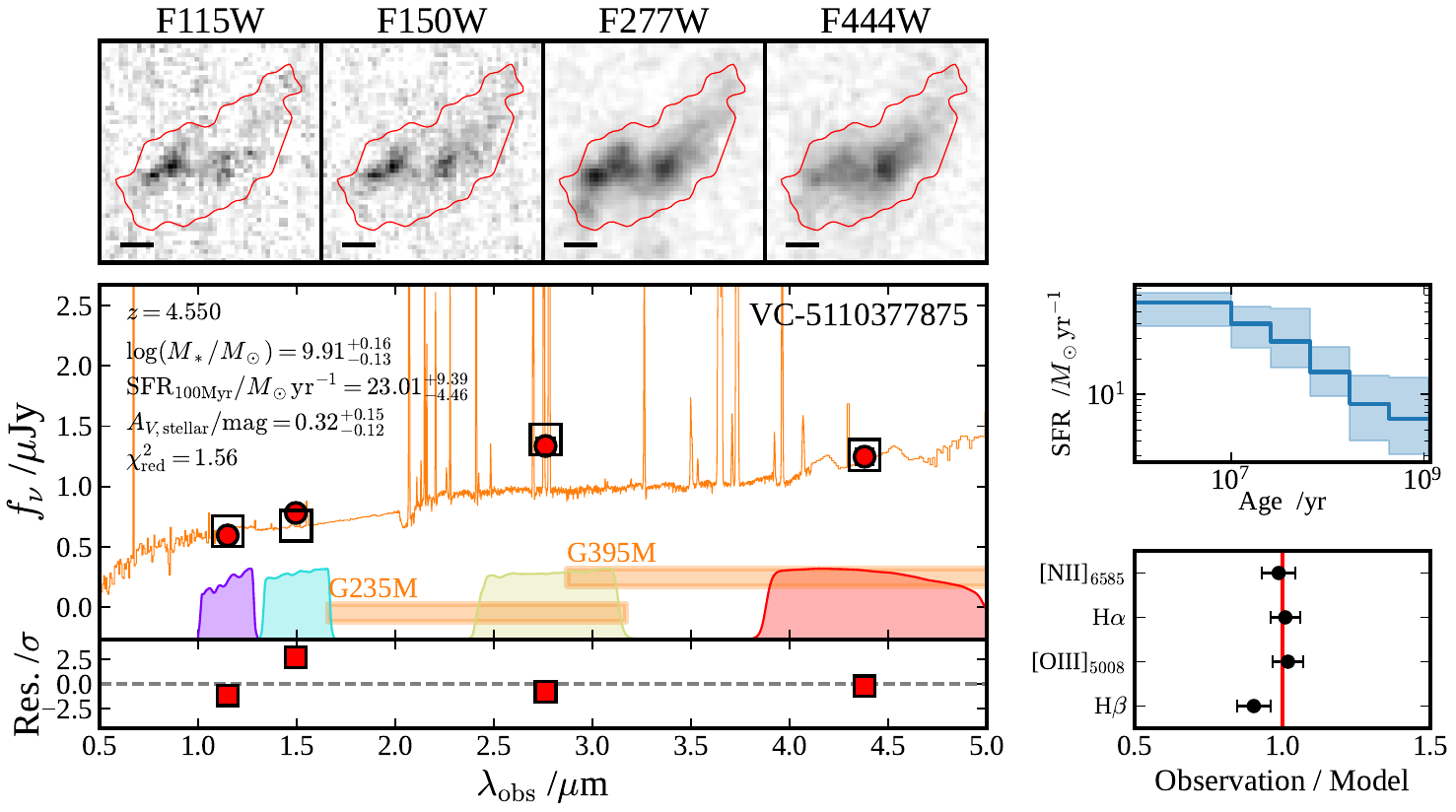}
\caption{An example of spatially-integrated SED fit for the same galaxy shown in Figure~\ref{fig:specfit}. (Top left) NIRCam images used for the SED fit, shown at their original resolution prior to PSF matching. The red lines show the mask to obtain photometry (see section \ref{sec:mask}). The scale bar at the lower left corner corresponds to $0.2\arcsec$.  (Bottom left) Best-fit SED, including observed photometry (red filled circles), model photometry (black open squares) and the model spectrum (orange line). Shaded curves represent the filter transmissions. Orange squares show the waverange coverages of the NIRSpec spectroscopy. The bottom part of the plot indicates the residual between the observed and modeled photometry, normalized by the flux uncertainties. (Right) Ratios of the observed line fluxes to the best-fit model predictions. The red vertical line marks the perfect match.  
\label{fig:sed}}
\end{figure*}

\begin{figure*}[!htbp]
\centering
\includegraphics[width=0.8\linewidth]{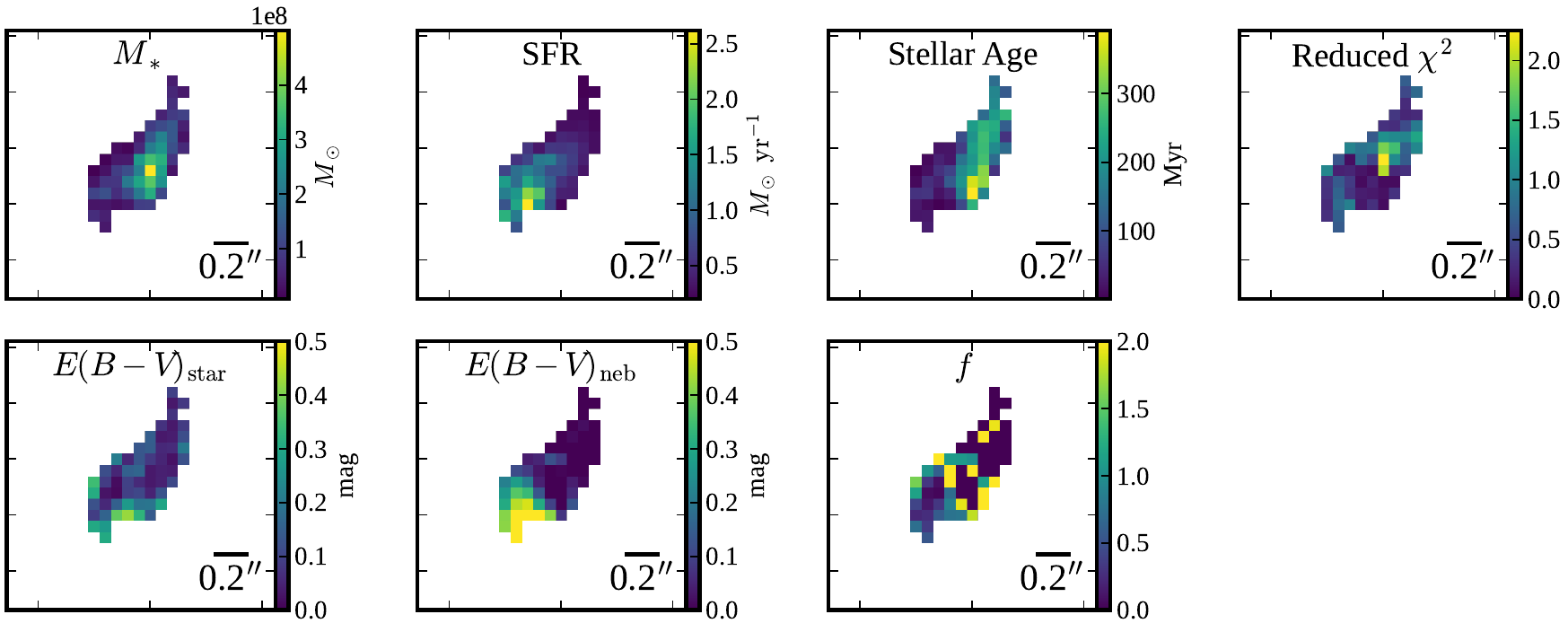}
\caption{Spatially resolved maps of the same galaxy shown in Figure~\ref{fig:sed}. 
Each panel shows a physical property derived from pixel-by-pixel SED fitting: 
stellar mass ($M_*$), SFR, mass-weighted stellar age, reduced $\chi^2$ of the SED fit, stellar reddening $\estar$, nebular reddening $\eneb$, 
and the differential reddening ratio $f=\estar/\eneb$. The scale bar at the lower left corner corresponds to $0.2\arcsec$.
\label{fig:map}}
\end{figure*}

The nebular emission (emission lines and continuum) is self-consistently modeled. We have two free parameters: the gas-phase metallicity ($Z_{\mathrm{gas}}$) and the ionization parameter ($U$). 
We assume a flat prior in log-space for the metallicity ($-2.0 < \log(Z_{\mathrm{gas}}/Z_\odot) < 0.5$) and ionization parameter ($-4 < \log(U) < -1$). 
We do not match the gas-phase metallicity $Z_{\mathrm{gas}}$ to the stellar metallicity $Z_{\mathrm{star}}$.
To account for systematic uncertainties in the underlying stellar population models, we always enforce a 5\% uncertainty floor on our flux measurements. The parameters and priors used in the SED fitting are summarized in Table~\ref{tbl:prospector_parameters}. 
Figure \ref{fig:sed} shows an example of spatially integrated SED fit for VC-5110377875 at $z=4.55$, highlighting the spatial region used for photometry and how the best-fit model reproduces the observed photometry and emission line fluxes.
Figure \ref{fig:map} shows an example of the spatially resolved maps of galaxy properties derived from pixel-by-pixel SED fitting for the same galaxy.
The integrated SED fitting results for the other galaxies in the sample are presented in Appendix~\ref{sec:appe_sed}.
A comparison of our SED fitting results with those from previous studies is presented in Appendix~\ref{appe:SEDfit_comparison}.


\section{Result and Discussion} \label{sec:result}
\subsection{Integrated and resolved physical properties} \label{sec:outshining}
We first discuss the physical properties derived from SED fitting for both spatially integrated and resolved (pixel-by-pixel) emission.
While SED fitting was performed for all galaxies in our sample, several objects are known to host AGN or show evidence for complex kinematics (e.g., outflows), which result in poor SED fits. 
To ensure a clean analysis, we exclude five such galaxies from the subsequent discussion: 
DC-417567a and DC-417567b (lack of NIRCam imaging), 
DC-536534 (identified as a robust broad-line AGN by \citealt{Ren2025arXiv250902027R}), 
DC-873321a (reported to host an obscured AGN by \citealt{Solimano2025A&A...693A..70S}), 
and DC-873756 (exhibiting complex velocity structures suggestive of outflows; \citealt{Ren2025arXiv250902027R}). 
The following analysis is therefore based on the remaining 15 galaxies.

The exceptional sensitivity and resolution of \jwst data have enabled spatially resolved studies of unlensed galaxies beyond cosmic noon (e.g., \citealt{Li2024ApJ...976...70L, Lines2025MNRAS.539.2685L}). Spatially resolving galaxies is crucial not only for understanding their local characteristics but also for improving the accuracy of global physical property estimates, such as total stellar mass derived from SED fitting. One of the key challenges in estimating $\mstar$ from spatially integrated (unresolved) emission is the so-called ``outshining'' effect \citep[e.g.,][]{Sorba2015MNRAS.452..235S, Clara2023ApJ...948..126G}. Essentially, young massive stars have significantly lower stellar mass-to-light ratios ($\mstar/L$) compared to older stars. Despite their lower mass contribution, young stars are orders of magnitude more luminous than solar-type stars and dominate the UV/optical SED of a galaxy. When the emission from an entire galaxy is integrated, the brighter contributions from young stars disproportionately influence the SED fit, often leading to an underestimation of the stellar mass contributed by the older, fainter population (e.g., \citealt{Sawicki1998AJ....115.1329S, Papovich2001ApJ...559..620P, Maraston2010MNRAS.407..830M, Clara2024A&A...686A..63G}). 
This effect could be particularly significant in high-z galaxies, where the ratio of light from young, star-forming regions relative to the underlying stellar mass is high and broad-band photometry is limited to the rest-frame UV/optical range ($\sim0.3–1.0$ dex at $z > 4$; e.g., \citealt{Clara2023ApJ...948..126G, Papovich2023ApJ...949L..18P, Narayanan2024ApJ...961...73N}).
To mitigate the potential bias, one can either conduct spatially resolved observations \citep[e.g.,][]{Sorba2015MNRAS.452..235S, Sorba2018MNRAS.476.1532S, Lines2025MNRAS.539.2685L} or, for unresolved data, adopt SFHs that account for contributions from recent starbursts, such as non-parametric SFHs \citep[e.g.,][]{Maraston2010MNRAS.407..830M, Pforr2012MNRAS.422.3285P, Leja2019ApJ...876....3L, Leja2019ApJ...877..140L, Carnall2024MNRAS.534..325C}, which enable more flexible modeling of the contributions from diverse stellar populations.
However, we note that \cite{Wuyts2012ApJ...753..114W} investigated massive star-forming galaxies at $z=0.5-2.5$ and showed that spatially resolved and integrated estimates of $\mstar$ generally agree well, while resolved SED modeling tends to yield older stellar ages compared to those inferred from integrated light.  
This suggests that discrepancies in total stellar mass estimates are not a universal effect but rather depend on the presence of non-uniform SFHs, the coexistence of regions with distinct colors within a galaxy, and the wavelength coverage of the observations.
For example, the lack of rest-frame near-infrared coverage--which is crucial for probing the light from older stellar populations--can exacerbate the outshining effect, particularly at high redshift, where observations are typically limited to rest-frame UV-to-optical wavelengths.

\begin{figure*}[!htbp]
\centering
\includegraphics[width=0.8\textwidth]{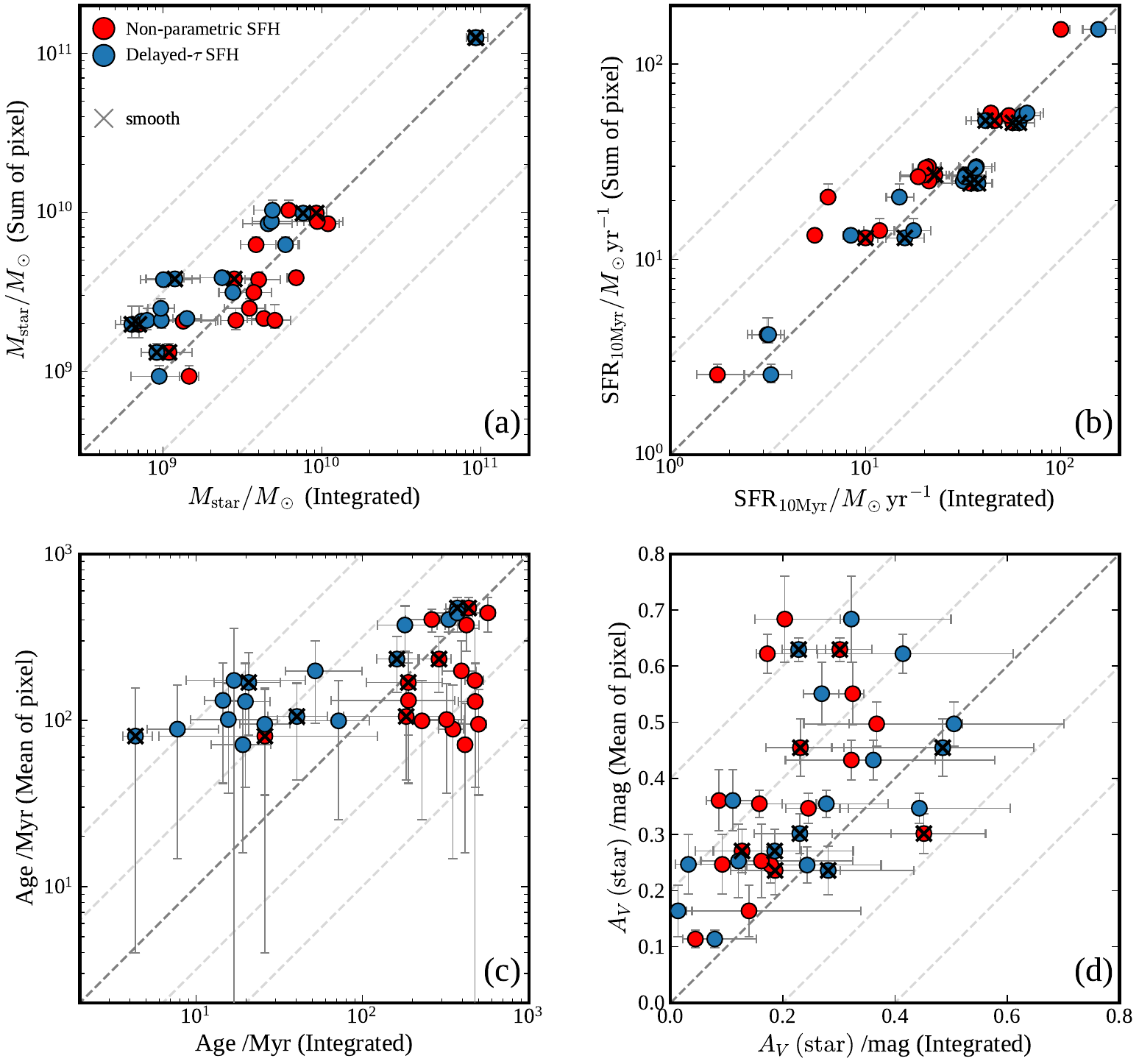}
\caption{Comparison of (a) total stellar mass, (b) SFR averaged over the last 10 Myr, (c) mass-weighted stellar age, and (d) $V$-band attenuation for the stellar continuum derived from spatially integrated SED fitting and pixel-by-pixel SED fitting. For spatially integrated SED fitting, results are shown for two SFH models: the non-parametric SFH (red points) and the parametric delayed-$\tau$ SFH (blue points). Symbols marked with a cross indicate galaxies that appear morphologically smooth (i.e., not clumpy or interacting) based on visual inspection. The thick dashed line represents the 1:1 relation, while the thin dashed lines indicate $\pm0.5$ dex and $\pm1$ dex offsets in (a)-(c) and $\pm0.2$ mag and $\pm0.4$ mag offsets in (d).
\label{fig:outshining}}
\end{figure*}

Figure \ref{fig:outshining} compares the total stellar mass ($M_{*}$), SFR averaged over the last 100 Myr (SFR$_{100\,\mathrm{Myr}}$), mass-weighted stellar age, and stellar attenuation ($A_V$) derived from SED fitting using spatially integrated emission with the results from pixel-by-pixel SED fitting for our sample. 
For $M_{*}$ and SFR$_{100\,\mathrm{Myr}}$, the pixel-by-pixel values are summed to obtain the total, and the uncertainties are computed by quadrature summation. 
The stellar age is the stellar-mass-weighted average of the pixel-by-pixel values, with the uncertainty given by the weighted standard deviation. $A_V$ based on the pixel-by-pixel measurements is computed by taking the ratio of the total V-band flux before and after attenuation, summed over all pixels.
For the spatially integrated SED fitting, we considered two approaches: one using the same parametric delayed-$\tau$ SFH as the pixel-by-pixel fits, and the other employing a non-parametric SFH with a continuity prior. 

Figure \ref{fig:outshining} (a) and (c) show a systematic trend where spatially integrated SED fitting using the parametric delayed-$\tau$ SFH yields lower stellar masses (median offset of $-0.26$ dex) and younger ages (median offset of $-0.57$ dex) compared to the pixel-by-pixel approach. This discrepancy is likely a consequence of the outshining effect, as the delayed-$\tau$ model may not fully capture localized starbursts, leading to an incomplete representation of the overall SFH of the galaxy.
In contrast, when using the non-parametric SFH for spatially integrated SED fitting, the results are consistent with the pixel-by-pixel ones in terms of stellar mass (median offset of $+0.03$ dex), but yield systematically older ages (median offset of $+0.27$ dex). 
A possible explanation for this discrepancy is that the non-parametric SFH may not sufficiently constrain the SFH, leading to results that are prior-dominated. The non-parametric SFHs are intentionally designed to include an early build-up of stellar mass and can be biased toward a more extended SFH and higher stellar masses \citep[e.g.,][]{Leja2019ApJ...876....3L, Leja2019ApJ...877..140L, Carnall2024MNRAS.534..325C}.
In contrast, as expected, the SFR$_{10\text{Myr}}$ estimates (Figure \ref{fig:outshining}b) show no significant systematic offset between spatially integrated and pixel-by-pixel approaches in either SFH case (+0.09 dex for the delayed-$\tau$ SFH and -0.11 dex for the non-parametric SFH). This behavior is expected because SFR$_{10\text{Myr}}$, which is sensitive to the young and bright stellar populations, is consistently recovered by both the integrated and resolved SED modeling. Using SFR$_{100\text{Myr}}$ instead yields similarly small offsets (-0.04 dex for delayed-$\tau$ and -0.12 dex for non-parametric SFH).

For the stellar attenuation (Figure \ref{fig:outshining}d), the uncertainties are large, but the pixel-by-pixel estimates are systematically higher than the integrated ones, regardless of the adopted SFH (median offset of $-0.16$ dex for the delayed-$\tau$ SFH and $-0.26$ dex for the non-parametric SFH). This systematic trend may reflect the patchy dust distribution within galaxies: integrated fits tend to be dominated by sightlines with lower attenuation, while pixel-based estimates capture the contribution of more obscured regions, leading to a larger effective $A_V$.
It is also worth noting that the parametric SFH adopted in the pixel-by-pixel SED fitting may be overly simplistic for actively star-forming regions. A more flexible representation of recent star formation variations could provide a better characterization of starburst/quenching \citep{Ciesla2017A&A...608A..41C}.

Finally, we also investigated whether the outshining effect depends on galaxy morphology.
As discussed above, outshining is expected to be particularly pronounced when galaxies host clumpy recent star formation or discontinuous activity triggered by interactions, situations in which a single SFH model cannot adequately reproduce the integrated SED.
To test this, we visually inspected the NIRCam images and separated the sample into galaxies that appear morphologically smooth and those showing clumpy or interacting features.
However, the stellar mass offset for the smooth subsample ($-0.16$ dex) is similar to that of the full sample, indicating no significant dependence on morphology.
The morphological classification (smooth or not) for each galaxy is indicated in the individual SED fitting figures in Appendix \ref{sec:appe_sed}.
All derived galaxy properties are summarized in Table \ref{tbl:sed_results}.


\subsection{Stellar versus nebular reddening}
\begin{figure}[htbp]
\centering
\includegraphics[width=\columnwidth]{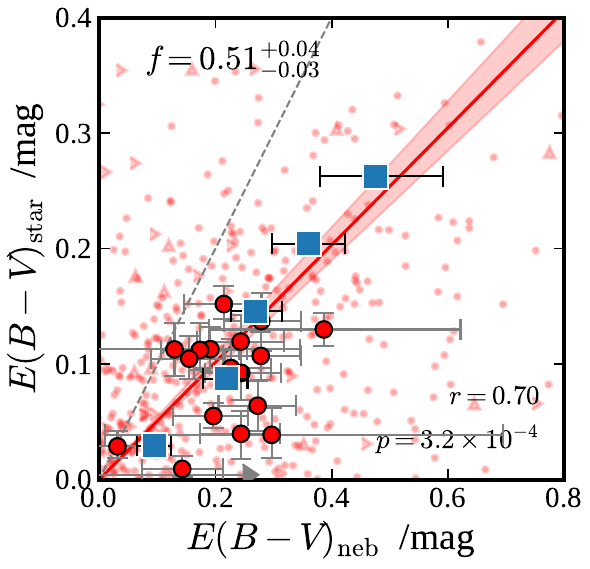}
\caption{Comparison of the stellar continuum and nebular line reddening for our sample. The red markers with black edges represent spatially integrated measurements, while the faint red circles and faint red triangles indicate pixel-by-pixel measurements (723 pixels). 
The circles correspond to reliable measurements, whereas the triangles indicate lower limits where H$\beta$ is undetected.  The blue squares are binned averages of the pixel-by-pixel measurements. The gray dotted line shows the 1:1 relation. The red solid line and shaded region indicate the best-fit linear relation obtained from the spatially integrated points together with the binned averages, and its associated uncertainty.}
\label{fig:f}
\end{figure}

Figure \ref{fig:f} shows the relation between $\estar$, derived from the SED fitting, and $\eneb$, measured from the Balmer decrement.
The figure includes both spatially integrated measurements obtained assuming the parametric SFH model (red points with black edges) and pixel-by-pixel results (faint red points; 723 pixels from the whole sample). 
For the pixel-by-pixel analysis, pixels where both \ha and \hb are undetected (S/N $<3$) are excluded. 
Additionally, blue squares indicate the binned averages of the pixel-by-pixel measurements.
Despite a large scatter, we find a positive correlation between stellar and nebular reddening: the Pearson correlation coefficient, computed by combining the spatially integrated and binned pixel-by-pixel results, is $r = 0.70$ with a $p$-value of $3.0 \times 10^{-4}$.
Fitting a linear relation to the combined data yields a best-fit ratio of $f = E(B{-}V)_\mathrm{star} / E(B{-}V)_\mathrm{neb} = 0.51^{+0.04}_{-0.03}$.
This value is slightly higher than the canonical ratio reported for local starburst galaxies ($f=0.44 \pm 0.03$; \citealt{Calzetti2000}).
Notably, using the non-parametric SFH model for the spatially integrated SED fitting yields a consistent result of $f = 0.48^{+0.03}_{-0.03}$.
However, due to the large scatter, this linear relation should be used with caution and is not necessarily applicable outside the range covered by the data.
We note that the above result is not significantly affected by uncertainties in the assumed intrinsic Balmer decrement $R_0$. 
While $R_0=2.86$ is typically adopted for Case~B recombination, \cite{Faisst2025survey} report higher electron temperatures for our sample, which would lower $R_0$ below 2.86. 
However, such variations change $\eneb$ by at most $\sim0.05$ mag, corresponding to a shift in $f$ of $< 10\%$ on median. 
Furthermore, Case~B itself may not strictly hold in some galaxies, as suggested by VC-5100541407b, which yields $E(B-V)_{\rm neb}<0$ (albeit consistent with zero within 1$\sigma$). 
Even adopting an extreme value of $R_0=2.62$ as reported by \citet{Scarlata2024arXiv240409015S}, the inferred $f$ would decrease by only $\sim10$\%. 

\begin{figure*}[htbp]
\centering
\includegraphics[width=\textwidth]{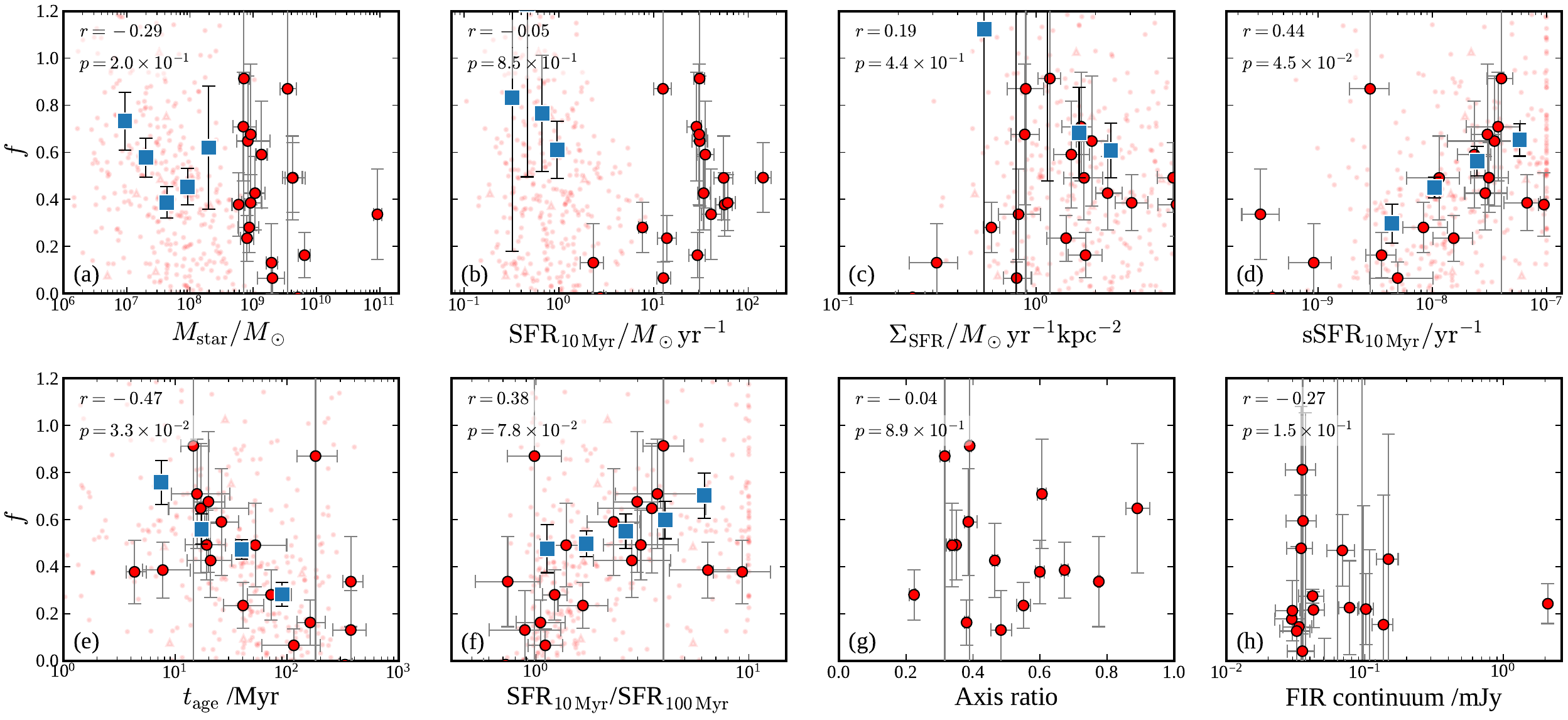}
\caption{
Plots of the derived dust reddening ratio as a function of various galaxy properties: (a) stellar mass, (b) SFR averaged over 10 Myr, (c) surface SFR density (based on SFR$_{\mathrm{10Myr}}$), (d) sSFR (based on SFR$_{\mathrm{10Myr}}$), (e) mass-weighted stellar age, (f) a burstiness proxy SFR$_{\mathrm{10Myr}}$/SFR$_{\mathrm{100Myr}}$, (g) axis ratio measured from the F444W image, and (h) FIR continuum flux. The red circles represent spatially integrated measurements, the faint red points indicate pixel-by-pixel results, and the blue squares show binned averages of the pixel-by-pixel data.
Axis ratio measurements are plotted only for sources that appear morphologically smooth. 
Pearson correlation coefficients ($r$) and $p$-values are calculated by combining the spatially integrated data points and the binned pixel-by-pixel results, and are shown in each panel.
\label{fig:correlation}}
\end{figure*}


We further investigate correlations between the $f$ factor and galaxy parameters in Figure \ref{fig:correlation}. 
In the two-component dust model (dense birth clouds and diffuse ISM), the $f$ factor mainly depends on the \hii-region volume filling factor and the density contrast between the two components (Appendix Eq.~\ref{eq:f_scaling}). 
These structural properties are expected to correlate with physical quantities such as stellar mass and SFR, motivating the examination of $f$ factor as a function of galaxy parameters.
The examined parameters include (a) total stellar mass, (b) SFR averaged over last 10 Myr, (c) surface SFR$_{\mathrm{10Myr}}$ density, (d) sSFR$_{\mathrm{10Myr}}={\rm SFR_{\mathrm{10Myr}}}/M_{\rm star}$, (e) mass-weighted stellar age, (f) a burstiness proxy defined as SFR$_{\mathrm{10Myr}}$/SFR$_{\mathrm{100Myr}}$, (g) the axis ratio measured in the  F444W image, and (h) FIR continuum flux. 
For the axis ratio, we consider only visually isolated sources in the F444W images and derive the median value along with the 16th and 84th percentile uncertainties from 1000 Monte Carlo fits.
For the FIR continuum flux, we examine positions within the ALMA maps that exhibit significant peak flux densities above $4\sigma$. Since the angular resolution of the ALMA data is coarser than that of the \jwst observations for several sources, we did not perform PSF matching between the two datasets. Instead, for each peak satisfying this threshold, we extracted the ALMA continuum flux and compared it with the average $f$ factor measured within the surrounding $2 \times 2$ pixel region. The associated uncertainties are estimated as the standard deviation among these measurements. When multiple $>4\sigma$ peaks are present in a single source, all such regions are included. 

The corresponding Pearson correlation coefficients (and $p$-values) are (a) -0.29 ($p=0.20$), (b) -0.05 ($p=0.85$), (c) 0.19 ($p=0.44$), (d) 0.44 ($p=0.045$), (e) -0.47 ($p=0.033$), (f) 0.38 ($p=0.078$), (g) -0.04 ($p=0.89$), and (h) -0.27 ($p=0.15$), respectively.
Among these, the strongest correlations in our sample are those with sSFR$_{10\,\mathrm{Myr}}$, stellar age, and the burstiness proxy SFR$_{\mathrm{10Myr}}$/SFR$_{\mathrm{100Myr}}$, though they remain statistically marginal given the limited sample size and scatter.
These correlations are consistent with a picture in which systems experiencing more recent, bursty star formation (and hence a larger fractional contribution from \hii regions) tend to exhibit higher $f$. In such systems, both nebular emission lines and stellar continuum are attenuated by a combination of diffuse dust and dense birth clouds, potentially leading to elevated $f$ factor values. Similar trends are observed in $z\sim1.5$ star-forming galaxies \citep[e.g.,][]{Price2014ApJ...788...86P}, and the physical picture is well illustrated in their Figure~5.

In contrast, we find no significant trend between $f$ and stellar mass or surface SFR density. Likewise, no significant correlation is observed with the axis ratio. At low redshift ($z_{\mathrm{median}} = 0.07$), \citet{Wild2011MNRAS.417.1760W} reported that galaxies with lower axis ratios (i.e., more edge-on) tend to exhibit elevated $f$ factors. This is interpreted as increased attenuation of the stellar continuum by diffuse dust along longer sightlines, while the attenuation of nebular emission from compact \hii regions remains relatively unaffected.
If this picture holds, the absence of such a correlation in our sample could imply that diffuse dust distribution is not disk-like but rather irregular, reducing inclination effects.
However, recent studies suggest that the physical picture at high redshift may differ from the classical two-component model. 
\citet{Reddy2020ApJ...902..123R} and \citet{Lorenz2023ApJ...951...29L} argue that in young, high-$z$ galaxies, both nebular lines and the UV continuum arise from young OB associations but along different sightlines: 
the nebular emission originates from the youngest, most embedded stars behind dustier columns, while the UV continuum is dominated by slightly older and less obscured stars. 
In this framework, $f$ factor reflects variations in dust covering fraction across sightlines rather than global disk geometry, and the lack of correlation with axis ratio in our sample can be understood in this context.

Figure~\ref{fig:correlation}(h) shows that at the positions of ALMA dust continuum peaks, the measured $f$ factors span the full range from $0$ to $1$ with no clear correlation with continuum flux. 
This may suggest that the bulk cold dust emission does not directly correspond to the additional attenuation affecting emission lines, which is likely caused by thick dust associated with star-forming regions. A clearer relation might emerge when probing finer physical scales or when comparing $f$ with total dust mass rather than continuum peak flux. 

\subsection{The Impact of the variation in reddening ratio}
As discussed in the Introduction, nebular attenuation is often inferred indirectly in high-redshift galaxies where emission line spectroscopy is unavailable or limited. 
A common approach is to estimate nebular reddening from stellar attenuation via an assumed differential reddening ratio $f$.
This assumption directly affects intrinsic line luminosities and derived quantities, such as the line luminosity function \citep[e.g.,][]{Saito2020MNRAS.494..199S, Covelo2024arXiv240917241C}, the H$\alpha$-based SFR (e.g., \citealt{faisst2022Univ....8..314F}), and the ionizing photon production efficiency $\xi_\mathrm{ion}$ \citep[e.g.,][]{Matthee2017MNRAS.465.3637M, Bouwens2016ApJ...831..176B}.
Here, we quantify how these quantities vary when adopting two different $f$ factors: our best-fit value of $f = 0.51$, which is slightly higher than the commonly used benchmark of $f = 0.44$ derived for local starburst galaxies \citep{Calzetti2000}, and the higher value of $f=1$, which is often assumed for high-z studies when the Balmer decrement is not available \citep[e.g.,][]{Asada2024ApJ...961..152A}.

The intrinsic \ha line luminosity ($F^{\text{intr}}_{\mathrm{H}\alpha}$) is corrected for dust attenuation based on the observed line luminosity ($F^{\mathrm{uncorr}}_{\mathrm{H}\alpha}$), with the correction depending on the assumed $f$-value as follows:
\begin{align*}
F^{\text{intr}}_{\mathrm{H}\alpha}&=F^{\mathrm{uncorr}}_{\mathrm{H}\alpha} \cdot 10^{0.4 A_{\mathrm{H}\alpha}}\\
&=F^{\mathrm{uncorr}}_{\mathrm{H}\alpha}\cdot10^{0.4 \frac{\estar}{f} k(\mathrm{H}\alpha)}.
\end{align*}

Based on the $E(B-V)_{\mathrm{star}}$ values of $0.009$–$0.15$ mag (median 0.10 mag) derived from our sample, and adopting $k(\mathrm{H}\alpha)$ from the \cite{Calzetti2000} attenuation curve with $R_V=4.05$, assuming $f=1$ leads to an underestimate of the intrinsic line luminosity by $0.01$–$0.19$~dex (median 0.13~dex), corresponding to $3-36$\% (median 25\%).
Consequently, this would lead to the line luminosity function and the CSFRD being systematically skewed toward lower values by the same amount.

The ionizing photon production efficiency $\xi_\mathrm{ion}$ is the amount of ionizing photons that are produced per unit UV luminosity and is defined as follows:
\begin{align*}
\xi^{\text{intr}}_\mathrm{ion} &= \frac{L^{\text{intr}}_{\mathrm{H}\beta}}{c_{\mathrm{H}\beta} L^{\text{intr}}_{\mathrm{UV}}} = \frac{L^{\text{uncorr}}_{\mathrm{H}\beta}}{c_{\mathrm{H}\beta}L^{\text{uncorr}}_{\mathrm{UV}}}\cdot \frac{10^{0.4A_\mathrm{\mathrm{H}\beta}}}{10^{0.4A_{\mathrm{UV}}}},\\
&=\xi^{\mathrm{uncorr}}_\mathrm{ion} \cdot 10^{0.4 \estar\left(\frac{k(\mathrm{H}\beta)}{f}-k(\mathrm{UV}) \right)}.
\end{align*}

Here, $L^{\text{intr}}_{\mathrm{H}\beta}$ and $L^{\text{intr}}_{\mathrm{UV}}$ are the intrinsic luminosities of H$\beta$ and the UV continuum, respectively, while $L^{\text{uncorr}}_{\mathrm{H}\beta}$ and $L^{\text{uncorr}}_{\mathrm{UV}}$ are their observed values. $c_{\mathrm{H}\beta}$ is the emission coefficient (e.g., \citealt{Matthee2023ApJ...950...67M}).
Under the same assumptions as above, assuming $f=1$ leads to an underestimate of the intrinsic $\xi_\mathrm{ion}$ by $0.05$–$0.25$~dex (median 0.13~dex), corresponding to a reduction of $13$–$78$\% (median 34\%).


\section{Summary} \label{sec:summary}
We analyzed 18 spectroscopically-confirmed star-forming galaxies at $z=4.4-5.7$ using \jwst/NIRSpec IFU and NIRCam data from the {\it ALPINE-CRISTAL JWST Survey}. By combining emission line fluxes with photometric data, we performed both pixel-by-pixel and spatially-integrated SED fitting with \pro, deriving the physical properties of the galaxies and evaluating their dust attenuation characteristics. Our main findings are as follows:
\begin{itemize}
    \item We investigated the impact of the outshining effect by comparing spatially integrated and pixel-by-pixel SED fitting. For spatially integrated fits using a parametric SFH, we found systematically lower stellar masses (median offset of -0.26 dex) and younger stellar ages (-0.57 dex) compared to pixel-based results, likely because integrated photometry is biased toward the light from a few bright young stars. 
    In contrast, integrated fits using a non-parametric SFH yielded older ages (+0.27 dex),  likely reflecting the model’s tendency to favor more extended SFH in the absence of strong constraints. 
    In terms of SFR, the integrated and pixel-based values showed good agreement, as expected given that both are dominated by recent star formation. 
    For stellar attenuation, the pixel-by-pixel estimates are systematically higher than the integrated ones, regardless of the adopted SFH, which may reflect the patchy dust distribution within galaxies.  These results underscore the importance of spatially resolved analyses and flexible SFH modeling in robustly recovering galaxy physical properties at high redshift.
    \item From the relationship between stellar and nebular dust attenuation, we derived a best-fit reddening ratio of $f = \estar/\eneb = 0.51^{+0.04}_{-0.03}$, slightly higher than the value of $f=0.44$ reported for local starburst galaxies by \citet{Calzetti2000}. This result provides a critical benchmark for dust attenuation correction in massive galaxies at $z>4$. 
    \item We explored how the $f$ factor varies with global galaxy properties. While the observed trends are modest, we find a positive correlations with sSFR and with a burstiness proxy SFR$_{\mathrm{10Myr}}$/SFR$_{\mathrm{100Myr}}$, and a negative correlation with stellar age, consistent with a larger fractional contribution from \hii regions in recently bursty systems, where both nebular lines and the stellar continuum are subject to attenuation by a combination of diffuse ISM and dense birth clouds.
    \item We quantified how the assumed $f$ factor affects key emission-line-derived quantities, such as the line luminosity function, line-based SFR, and ionizing photon production efficiency $\xi_\mathrm{ion}$. Assuming $f = 1$ instead of our derived value of $f = 0.51$ would lead to an underestimation of the intrinsic line luminosity by $0.01$--$0.19$~dex (median $0.13$~dex; $3$--$36$\%) and the ionizing photon production efficiency $\xi_\mathrm{ion}$ by $0.05$--$0.25$~dex (median $0.13$~dex; $13$--$78$\%). 
\end{itemize}
Our findings emphasize the importance of accurately measuring $f$ for robust dust attenuation corrections, particularly in high-redshift studies. 

\begin{acknowledgments}
We thank the anonymous referee for constructive comments that improved the manuscript.
The JWST and HST data presented in this article were obtained from the Mikulski Archive for Space Telescopes (MAST) at the Space Telescope Science Institute. 
These observations are associated with programs, HST-GO-13641 (\dataset[doi: 10.17909/xne1-7v26]{https://doi.org/10.17909/xne1-7v26}), JWST-GO-01727 (\dataset[doi: 10.17909/ph8h-qf05]{https://doi.org/10.17909/ph8h-qf05}), JWST-GO-03045 (\dataset[doi: 10.17909/cqds-qc81]{https://doi.org/10.17909/cqds-qc81}), and JWST-GO-04265 (\dataset[doi:10.17909/wac6-9741]{https://doi.org/10.17909/wac6-9741}).
Data analysis was in part carried out on the Multi-wavelength Data Analysis System operated by the Astronomy Data Center (ADC), NAOJ. This research was supported by FoPM, WINGS Program, the University of Tokyo. A.T. acknowledges the support by JSPS KAKENHI Grant Number 24KJ0562.
M.B. gratefully acknowledges support from the ANID BASAL project FB210003. This work was supported by the French government through the France 2030 investment plan managed by the National Research Agency (ANR), as part of the Initiative of Excellence of Université Côte d'Azur under reference number ANR-15-IDEX-01.
A.N. acknowledges support from the Narodowe Centrum Nauki (NCN), Poland, through the SONATA BIS grant UMO2020/38/E/ST9/00077. 
P.S. acknowledges support from the Narodowe Centrum Nauki (NCN), Poland, through the SONATA BIS grant UMO2020/38/E/ST9/00077.
V.V. acknowledges support from the ALMA-ANID Postdoctoral Fellowship under the award ASTRO21-0062.
M.A. acknowledges support from ANID Basal Project FB210003 and and ANID MILENIO NCN2024\_112.
E.~I. acknowledge funding by ANID FONDECYT Regular 1221846 and also financial support from ANID - MILENIO - NCN2024\_112.
I.D.L. acknowledges funding from the European Research Council (ERC) under the European Union's Horizon 2020 research and innovation program DustOrigin (ERC-2019- StG-851622), from the Belgian Science Policy Office (BELSPO) through the PRODEX project “JWST/MIRI Science exploitation” (C4000142239) and from the Flemish Fund for Scientific Research (FWO-Vlaanderen) through the research project G0A1523N.
J.~M. acknowledge financial support from ANID - MILENIO - NCN2024\_112.
K.~K. acknowledges the support by JSPS KAKENHI Grant Numbers JP22H04939, JP23K20035, and JP24H00004.
\end{acknowledgments}

\facilities{\hst(WFC3), \jwst(NIRCam and NIRSpec), ALMA}

\software{Astropy \citep{astropy:2022},
          Photutil \citep{larry_bradley_2024_13989456},
          Numpy \citep{harris2020array},
          Prospector \citep{Leja2017ApJ...837..170L, Johnson2021ApJS..254...22J}, 
          Dynesty \citep{Speagle2020}
          }



\appendix
\section{A simple model for the relation between $f$ and galaxy properties} \label{sec:appe_f}
We outline a simple two-component (diffuse ISM and dense \hii birth clouds) model to build intuition for how the $f$ factor relates to galaxy properties.
Consider a galaxy whose diffuse ISM (diffuse dust) has mean gas density $n$, within which dense \hii regions of gas density $n_i$ are randomly distributed. 
Let $r_e$ denote the line-of-sight path length through the diffuse component, and $r_i$ the cumulative line-of-sight path length spent inside the birth clouds along the same sightline.
Assuming spatially uniform dust properties, the effective optical depths for the stellar continuum ($\tau_{\text{cont}}$) and for nebular emission ($\tau_{\text{neb}}$) can be written as
\[
\tau_{\text{cont}}=n\,\sigma_{\text{cont}}\,r_e,
\qquad
\tau_{\rm neb}=n_i\,\sigma_{\rm neb}\,r_i+n\,\sigma_{\rm neb}\,(r_e-r_i),
\]
where $\sigma_\lambda$ denotes the dust extinction cross section per particle.
Taking the ratio gives
\[
\frac{\tau_{\rm neb}}{\tau_{\text{cont}}}
= \frac{\sigma_{\rm neb}}{\sigma_{\text{cont}}}
\left[\,1+\left(\frac{r_i}{r_e}\right)\delta n\,\right],
\]
with $\delta n \equiv (n_i-n)/n$ the density contrast between the birth clouds and the diffuse ISM.
We adopt the same curves as in the main text—\citet{Calzetti2000} for the stellar continuum and a MW extinction curve \citep{Cardelli1989ApJ...345..245C} for nebular lines.
Using $k_\lambda \equiv A_\lambda/E(B\!-\!V)$ and $A_\lambda=1.086\,\tau_\lambda$, we write
\[
\frac{1}{f}
= \frac{k^{\text{Cal}}_{\text{cont}}}{k^{\text{MW}}_{\rm neb}}\frac{\tau_{\rm neb}}{\tau_{\text{cont}}} 
= \frac{k^{\text{Cal}}_{\text{cont}}}{k^{\text{MW}}_{\rm neb}}\frac{\sigma_{\text{neb}}}{\sigma_{\text{cont}}}
\left[1+\left(\frac{r_i}{r_e}\right)\delta n\right].
\]
Here, $k^{\text{Cal}}_{\text{cont}}$ is the attenuation–curve value for the stellar continuum under the Calzetti attenuation law, and $k^{\text{MW}}_{\rm neb}$ is the extinction–curve value for nebular lines under the MW extinction law.
As the cross-section ratio is written as the ratio of extinction-curve values (extinction laws assume a foreground-screen geometry), we can write
\begin{equation}
\frac{1}{f}
= \frac{k^{\text{Cal}}_{\text{cont}}}{k^{\text{MW}}_{\text{cont}}}
\left[\,1+\left(\frac{r_i}{r_e}\right)\delta n\right]
=
1.25\,\left[\,1+\left(\frac{r_i}{r_e}\right)\delta n\right].
\label{eq:f_scaling}
\end{equation}
Here we take the continuum at $1500\,\text{\AA}$.
This equation shows that $f$ factor depends on how much of the line of sight passes through star-forming regions and on the density contrast between the birth clouds and the diffuse ISM.
The $f$ factor depends on physical properties such as the SFR, gas density, and size, through the quantities $r_i/r_e$ and $\delta n$.

As a fiducial example, we insert MW values into Eq.~\eqref{eq:f_scaling}.
Assuming the \hii regions are randomly distributed in the disk, the line-of-sight path–length fraction equals the \hii volume filling factor, so we set $r_i/r_e \simeq 5\times10^{-3}$ (for the MW; \citealt{Berkhuijsen1998LNP...506..301B}).
For the density contrast, we take a diffuse-ISM number density of $\sim 1\,\mathrm{cm^{-3}}$ and a GMC mean density of $\sim 100\,\mathrm{cm^{-3}}$ \citep[e.g.,][]{Ferriere2001RvMP...73.1031F}, which implies $\delta n\sim100$.
Substituting these values into Eq.~\eqref{eq:f_scaling} yields $f\sim0.5$.
Although highly simplified—especially in assuming that the stellar continuum does not pass through dense star-forming regions, this estimate is of the correct order of magnitude and captures the essential dependence.



\section{SED fitting} \label{sec:appe_sed}
Table \ref{tbl:prospector_parameters} shows the parameters and their priors used in the \pro SED model. Table \ref{tbl:sed_results} shows the obtained physical parameters including \ha-based spectroscopic redshift, total stellar mass, SFR averaged over last 100 Myr, dust attenuation for stellar continuum, and dust attenuation for nebular lines. 

\begin{deluxetable}{lll}[!hbt]
\tablewidth{0.43\textwidth}
\tablecaption{Summary of parameters and priors used in the SED model \label{tbl:prospector_parameters}}
\tablehead{
\colhead{Parameter} & \colhead{Prior Functions}
}
\startdata
\cutinhead{SFH (parametric)}
$\log (M_\star/M_{\odot})$             & Uniform(5, 12)   \\
$\log (\tau /\mathrm{Gyr}^{-1})$                        & Uniform(-1, 1.5)   \\
$\log (t_\mathrm{age}/\mathrm{Gyr})$                    & Uniform(-2, $t(z)$\tablenotemark{a})  \\
$\log(Z_\star/Z_\odot)$                                 & Uniform(-2.0, 0.2) \\
\cutinhead{SFH (non-parametric)}
$\log (M_\star/M_{\odot})$             & Uniform(8, 12)   \\
Number of bins                                          & 6  \\
$\log r_i$\tablenotemark{b}                             & StudentT(0, 0.3) \\
$\log(Z_\star/Z_\odot)$                                 & Uniform(-2.0, 0.2) \\
\cutinhead{Dust Attenuation}
Attenuation curve                                       & \cite{Calzetti2000} \\
$A_{V, \mathrm{star}}$                                  & Uniform(0, 4.0) \\
$\delta$                                                & Uniform(-1.0, 0.4) \\
\cutinhead{Nebular Emission}
$\log (Z_{\rm gas}/Z_\odot)$                            & Uniform(-2, 0.5) \\
$\log U_{\rm neb}$                                      & Uniform(-4, -1) \\
\enddata
\tablecomments{Uniform($x$,$y$) indicates a uniform distribution from $x$ to $y$, while StudentT($\mu$, $\sigma$) indicates a Student's $t$-distribution with mean of $\mu$ and size of $\sigma$.}
\tablenotetext{a}{$t(z)$ is the cosmic age at redshift $z$.}
\tablenotetext{b}{$r_i$ is the ratio of the SFR in temporal bin $i$ to that in the adjacent bin. There are 5 such parameters that describe the SFH.}
\end{deluxetable}

\movetabledown=60mm
\begin{rotatetable}
\begin{deluxetable*}{lccccccccccccc}
  \tabletypesize{\footnotesize}
\tablecolumns{14}
\tablewidth{6in}
\tablecaption{Derived physical parameters.} \label{tbl:sed_results}
\tablehead{
  \colhead{} & \colhead{} & \colhead{} & \colhead{} &
  \multicolumn{3}{c}{$\log (M_\ast / M_\odot$)} &
  \multicolumn{3}{c}{$\mathrm{SFR} / M_\odot~\mathrm{yr}^{-1}$} &
  \multicolumn{3}{c}{$A_{V, \mathrm{star}}$} &
  \colhead{$A_{V, \mathrm{neb}}$} \\[-6pt]
  \colhead{ID} & \colhead{R.A.} & \colhead{Dec.} &
    \colhead{$z_{\mathrm{H}\alpha}$} &
    \colhead{Para}  & \colhead{Non-Para}  & \colhead{Sum} &
    \colhead{Para}  & \colhead{Non-Para}  & \colhead{Sum} &
    \colhead{Para}  & \colhead{Non-Para}  & \colhead{Mean} &
    \colhead{}\\
  \colhead{(1)} & \colhead{(2)} & \colhead{(3)} &
    \colhead{(4)} & \colhead{(5)} & \colhead{(6)} & \colhead{(7)} & \colhead{(8)} & \colhead{(9)} & \colhead{(10)} & \colhead{(11)} & \colhead{(12)} & \colhead{(13)} & \colhead{(14)}
}
\startdata
DC-417567a/C-10a\dag & 10:02:04.08 & 01:55:44.15 & 5.666 & $8.87_{-0.08}^{+0.09}$ & $9.89_{-0.04}^{+0.04}$ & $8.95_{-0.05}^{+0.08}$ & $7.76_{-1.33}^{+1.83}$ & $33.62_{-6.25}^{+6.05}$ & $9.26_{-1.06}^{+1.78}$ & $0.14_{-0.02}^{+0.04}$ & $0.47_{-0.04}^{+0.05}$ & $0.53_{-0.10}^{+0.10}$ & $0.42_{-0.23}^{+0.65}$ \\
DC-417567b/C-10a\dag & 10:02:04.14 & 01:55:44.50 & 5.671 & $8.76_{-0.18}^{+0.20}$ & $9.38_{-0.12}^{+0.18}$ & $8.83_{-0.10}^{+0.16}$ & $6.01_{-2.04}^{+3.53}$ & $5.16_{-2.11}^{+4.59}$ & $6.98_{-1.50}^{+2.78}$ & $0.76_{-0.21}^{+0.25}$ & $0.63_{-0.16}^{+0.28}$ & $0.78_{-0.21}^{+0.21}$ & $0.70_{-0.49}^{+0.89}$ \\
DC-494763/C-19 & 10:00:05.10 & 02:03:12.20 & 5.233 & $9.07_{-0.18}^{+0.16}$ & $9.45_{-0.08}^{+0.11}$ & $9.58_{-0.04}^{+0.04}$ & $13.35_{-4.56}^{+5.77}$ & $15.02_{-1.59}^{+1.39}$ & $17.87_{-0.88}^{+0.87}$ & $0.23_{-0.08}^{+0.16}$ & $0.45_{-0.16}^{+0.11}$ & $0.30_{-0.04}^{+0.04}$ & $0.70_{-0.50}^{+0.91}$ \\
DC-519281/C-09a & 09:59:00.89 & 02:05:27.65 & 5.575 & $8.80_{-0.10}^{+0.10}$ & $8.85_{-0.07}^{+0.16}$ & $9.29_{-0.09}^{+0.11}$ & $6.57_{-1.40}^{+1.67}$ & $6.68_{-0.86}^{+1.22}$ & $15.60_{-2.65}^{+1.60}$ & $0.48_{-0.20}^{+0.16}$ & $0.23_{-0.06}^{+0.08}$ & $0.46_{-0.05}^{+0.05}$ & $0.77_{-0.56}^{+0.98}$ \\
DC-536534/C-03\dag & 09:59:53.25 & 02:07:05.41 & 5.689 & $9.26_{-0.10}^{+0.10}$ & $9.51_{-0.17}^{+0.21}$ & $9.79_{-0.06}^{+0.08}$ & $18.69_{-4.00}^{+4.79}$ & $24.71_{-5.27}^{+5.51}$ & $60.07_{-7.00}^{+8.72}$ & $1.23_{-0.22}^{+0.17}$ & $0.79_{-0.10}^{+0.09}$ & $1.72_{-0.08}^{+0.08}$ & $0.88_{-0.67}^{+1.09}$ \\
DC-630594/C-11a & 10:00:32.59 & 02:15:28.46 & 4.440 & $9.15_{-0.09}^{+0.09}$ & $9.63_{-0.10}^{+0.08}$ & $9.33_{-0.02}^{+0.03}$ & $15.95_{-2.93}^{+3.23}$ & $7.81_{-1.49}^{+2.46}$ & $15.69_{-0.52}^{+0.66}$ & $0.44_{-0.14}^{+0.16}$ & $0.25_{-0.04}^{+0.07}$ & $0.35_{-0.03}^{+0.03}$ & $0.60_{-0.39}^{+0.81}$ \\
DC-683613/C-05 & 10:00:09.42 & 02:20:13.90 & 5.536 & $9.88_{-0.10}^{+0.09}$ & $9.96_{-0.06}^{+0.07}$ & $9.99_{-0.02}^{+0.03}$ & $32.60_{-5.33}^{+6.20}$ & $26.21_{-4.33}^{+4.86}$ & $27.66_{-1.09}^{+1.85}$ & $0.19_{-0.11}^{+0.14}$ & $0.13_{-0.08}^{+0.11}$ & $0.27_{-0.04}^{+0.04}$ & $0.78_{-0.56}^{+0.96}$ \\
DC-709575/C-14 & 09:59:47.07 & 02:22:32.95 & 4.412 & $8.96_{-0.10}^{+0.10}$ & $9.04_{-0.11}^{+0.14}$ & $9.12_{-0.03}^{+0.06}$ & $9.55_{-1.35}^{+1.49}$ & $6.48_{-0.89}^{+1.06}$ & $8.88_{-0.41}^{+0.62}$ & $0.28_{-0.15}^{+0.15}$ & $0.19_{-0.08}^{+0.12}$ & $0.24_{-0.04}^{+0.04}$ & $0.84_{-0.64}^{+1.04}$ \\
DC-742174/C-17 & 10:00:39.12 & 02:25:32.20 & 5.636 & $8.98_{-0.18}^{+0.15}$ & $9.16_{-0.06}^{+0.06}$ & $8.97_{-0.04}^{+0.07}$ & $6.86_{-0.68}^{+0.53}$ & $5.97_{-0.82}^{+0.89}$ & $6.64_{-0.52}^{+0.81}$ & $0.11_{-0.03}^{+0.04}$ & $0.09_{-0.02}^{+0.03}$ & $0.36_{-0.05}^{+0.05}$ & $0.59_{-0.40}^{+0.79}$ \\
DC-842313a/C-01a & 10:00:54.49 & 02:34:36.03 & 4.544 & $9.79_{-0.09}^{+0.11}$ & $10.58_{-0.12}^{+0.08}$ & $10.51_{-0.04}^{+0.08}$ & $64.32_{-12.26}^{+18.17}$ & $114.49_{-25.13}^{+48.92}$ & $122.38_{-14.04}^{+29.48}$ & $1.99_{-0.26}^{+0.29}$ & $2.05_{-0.11}^{+0.09}$ & $1.79_{-0.08}^{+0.08}$ & $2.78_{-2.53}^{+3.06}$ \\
DC-842313b/C-01a & 10:00:54.53 & 02:34:34.62 & 4.552 & $9.37_{-0.24}^{+0.19}$ & $9.84_{-0.06}^{+0.04}$ & $9.59_{-0.03}^{+0.04}$ & $13.46_{-1.31}^{+1.24}$ & $15.02_{-2.90}^{+3.94}$ & $14.78_{-0.96}^{+1.63}$ & $0.03_{-0.02}^{+0.06}$ & $0.09_{-0.07}^{+0.08}$ & $0.25_{-0.05}^{+0.05}$ & $0.45_{-0.23}^{+0.68}$ \\
DC-848185/C-02 & 10:00:21.51 & 02:35:10.91 & 5.294 & $9.66_{-0.16}^{+0.15}$ & $10.04_{-0.08}^{+0.07}$ & $9.93_{-0.03}^{+0.04}$ & $50.22_{-15.56}^{+20.58}$ & $23.31_{-3.56}^{+3.32}$ & $70.74_{-3.11}^{+4.00}$ & $0.41_{-0.15}^{+0.20}$ & $0.17_{-0.02}^{+0.04}$ & $0.62_{-0.04}^{+0.04}$ & $0.85_{-0.67}^{+1.06}$ \\
DC-873321a/C-07a\dag & 10:00:04.06 & 02:37:35.87 & 5.154 & $9.00_{-0.14}^{+0.18}$ & $9.60_{-0.09}^{+0.13}$ & $9.58_{-0.03}^{+0.04}$ & $10.47_{-2.97}^{+5.32}$ & $14.29_{-2.27}^{+3.65}$ & $29.67_{-1.75}^{+2.04}$ & $0.51_{-0.19}^{+0.20}$ & $0.37_{-0.13}^{+0.15}$ & $0.50_{-0.04}^{+0.04}$ & $0.88_{-0.66}^{+1.07}$ \\
DC-873321b/C-07b & 10:00:03.97 & 02:37:36.36 & 5.154 & $8.90_{-0.17}^{+0.20}$ & $9.46_{-0.10}^{+0.23}$ & $9.32_{-0.06}^{+0.08}$ & $8.73_{-2.78}^{+5.09}$ & $7.07_{-2.15}^{+2.95}$ & $13.86_{-0.92}^{+1.37}$ & $0.27_{-0.03}^{+0.06}$ & $0.33_{-0.03}^{+0.02}$ & $0.55_{-0.06}^{+0.06}$ & $0.67_{-0.46}^{+0.88}$ \\
DC-873756/C-24\dag & 10:00:02.71 & 02:37:40.14 & 4.544 & $10.11_{-0.18}^{+0.34}$ & $10.61_{-0.26}^{+0.25}$ & $10.46_{-0.04}^{+0.09}$ & $83.11_{-14.18}^{+21.43}$ & $57.23_{-20.26}^{+49.63}$ & $89.85_{-6.39}^{+13.93}$ & $1.28_{-0.23}^{+0.33}$ & $1.45_{-0.35}^{+0.46}$ & $1.75_{-0.07}^{+0.07}$ & $1.78_{-1.78}^{+1.78}$ \\
VC-5100541407a/C-06a & 10:01:00.91 & 01:48:33.67 & 4.563 & $9.69_{-0.12}^{+0.14}$ & $9.79_{-0.07}^{+0.09}$ & $10.01_{-0.04}^{+0.06}$ & $17.89_{-3.18}^{+3.91}$ & $12.06_{-2.05}^{+2.02}$ & $14.24_{-1.04}^{+2.04}$ & $0.32_{-0.11}^{+0.18}$ & $0.20_{-0.05}^{+0.10}$ & $0.68_{-0.08}^{+0.08}$ & $0.42_{-0.14}^{+1.02}$ \\
VC-5100541407b/C-06b & 10:01:00.99 & 01:48:35.00 & 4.562 & $9.77_{-0.06}^{+0.08}$ & $9.59_{-0.10}^{+0.27}$ & $9.80_{-0.03}^{+0.05}$ & $4.42_{-0.75}^{+0.67}$ & $11.05_{-2.13}^{+2.92}$ & $4.87_{-0.39}^{+0.93}$ & $0.01_{-0.01}^{+0.03}$ & $0.14_{-0.11}^{+0.20}$ & $0.16_{-0.05}^{+0.05}$ & $-0.95_{-0.95}^{+0.95}$ \\
VC-5100822662a/C-04a & 09:58:57.90 & 02:04:51.41 & 4.520 & $8.99_{-0.18}^{+0.35}$ & $9.70_{-0.14}^{+0.10}$ & $9.32_{-0.05}^{+0.10}$ & $10.48_{-3.49}^{+10.82}$ & $9.67_{-2.25}^{+1.91}$ & $12.42_{-0.98}^{+1.04}$ & $0.24_{-0.07}^{+0.13}$ & $0.18_{-0.04}^{+0.07}$ & $0.24_{-0.03}^{+0.03}$ & $0.54_{-0.33}^{+0.74}$ \\
VC-5100822662b/C-04b & 09:58:57.94 & 02:04:53.00 & 4.521 & $9.44_{-0.09}^{+0.09}$ & $9.57_{-0.10}^{+0.11}$ & $9.50_{-0.04}^{+0.05}$ & $3.69_{-0.94}^{+0.94}$ & $2.32_{-0.89}^{+1.05}$ & $2.86_{-0.24}^{+0.34}$ & $0.12_{-0.07}^{+0.11}$ & $0.16_{-0.11}^{+0.16}$ & $0.25_{-0.07}^{+0.07}$ & $0.94_{-0.11}^{+2.08}$ \\
VC-5100994794/C-13a & 10:00:41.17 & 02:17:14.29 & 4.580 & $8.99_{-0.08}^{+0.09}$ & $9.54_{-0.16}^{+0.10}$ & $9.40_{-0.02}^{+0.06}$ & $10.82_{-1.84}^{+2.39}$ & $7.75_{-1.03}^{+1.70}$ & $14.90_{-0.58}^{+0.69}$ & $0.28_{-0.08}^{+0.11}$ & $0.16_{-0.04}^{+0.10}$ & $0.35_{-0.03}^{+0.03}$ & $0.48_{-0.28}^{+0.67}$ \\
VC-5101218326/C-25 & 10:01:12.51 & 02:18:52.69 & 4.573 & $10.97_{-0.06}^{+0.07}$ & $10.97_{-0.04}^{+0.05}$ & $11.10_{-0.01}^{+0.01}$ & $55.91_{-12.00}^{+16.54}$ & $158.44_{-24.16}^{+27.56}$ & $60.51_{-1.88}^{+2.93}$ & $0.23_{-0.03}^{+0.05}$ & $0.30_{-0.04}^{+0.06}$ & $0.63_{-0.02}^{+0.02}$ & $1.19_{-0.55}^{+1.90}$ \\
VC-5101244930/C-15 & 10:00:47.66 & 02:18:02.09 & 4.581 & $8.87_{-0.07}^{+0.08}$ & $9.13_{-0.17}^{+0.20}$ & $9.31_{-0.03}^{+0.04}$ & $7.98_{-1.22}^{+1.68}$ & $7.44_{-1.04}^{+1.21}$ & $12.63_{-0.41}^{+0.42}$ & $0.08_{-0.04}^{+0.07}$ & $0.04_{-0.02}^{+0.04}$ & $0.11_{-0.02}^{+0.02}$ & $0.08_{-0.12}^{+0.28}$ \\
VC-5110377875 & 10:01:32.34 & 02:24:30.32 & 4.550 & $9.68_{-0.12}^{+0.20}$ & $9.97_{-0.13}^{+0.16}$ & $9.94_{-0.02}^{+0.04}$ & $45.84_{-6.51}^{+8.86}$ & $26.53_{-5.15}^{+10.82}$ & $35.54_{-1.49}^{+2.51}$ & $0.36_{-0.13}^{+0.22}$ & $0.32_{-0.12}^{+0.15}$ & $0.43_{-0.04}^{+0.04}$ & $0.77_{-0.51}^{+1.01}$ \\
\enddata
%
%
\tablecomments{(1) Galaxy ID, with the CRISTAL ID shown after the slash. The suffixes ``a'' and ``b'' indicate components of a galaxy system, such as merger pairs.\\
(2)--(4) right ascension, declination, \ha-based spectroscopic redshift of target\\
(5-6) stellar mass estimated by SED fitting using spatially integrated emission with a parametric and non-parametric SFH\\
(7) sum of stellar mass estimated by pixel-by-pixel SED fitting\\
(8-9) SFR estimated by spatially integrated SED fitting (all SFRs are averaged over the last 100 Myr) with a parametric and non-parametric SFH\\
(10) sum of SFR estimated by pixel-by-pixel SED fitting\\
(11-12) V-band attenuation of stellar continuum measured with the SED fitting with a parametric anf non-parametric SFH\\
(13) Mean of V-band attenuation of stellar continuum estimated by pixel-by-pixel SED fitting\\
(14) V-band attenuation of nebular lines measured with the Balmer decrement\\
\noindent\dag~Galaxies marked with $\dag$ are excluded from the final analysis. For the reasons for exclusion, see Section~\ref{sec:outshining}.}
\end{deluxetable*}
\end{rotatetable}

\begin{figure*}[htbp]
\centering
\includegraphics[width=0.9\textwidth]{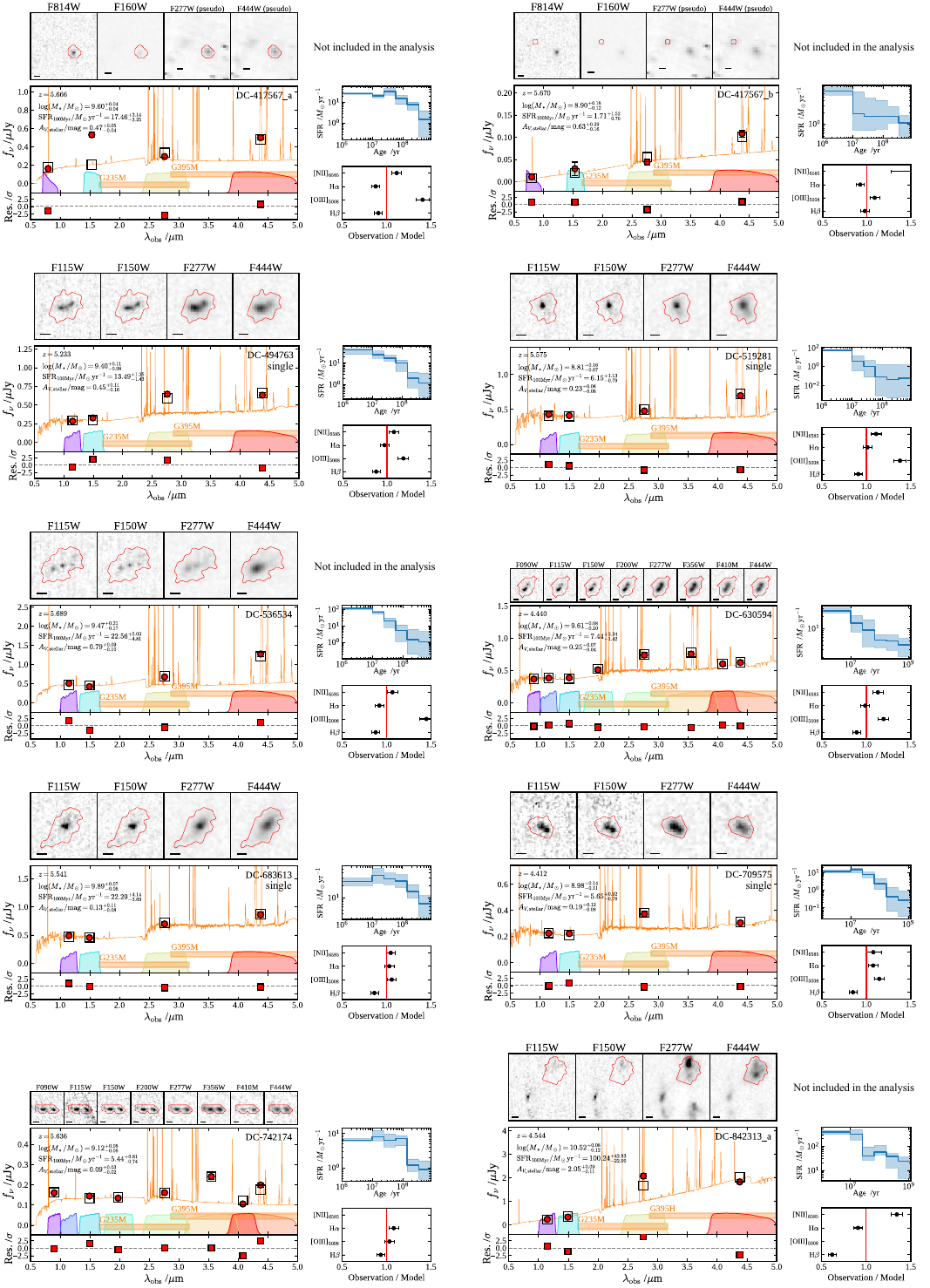}
\caption{Same as Figure~\ref{fig:sed}, but for all galaxies in our sample.
Objects labeled with ``single’’ below their IDs are those visually classified as smooth sources (i.e., neither interacting nor clumpy).
These correspond to the galaxies marked with crosses in Figure~\ref{fig:outshining}.
\label{fig:SED_Appe_1}}
\end{figure*}
\begin{figure*}[!h]
\centering
\includegraphics[width=0.9\textwidth]{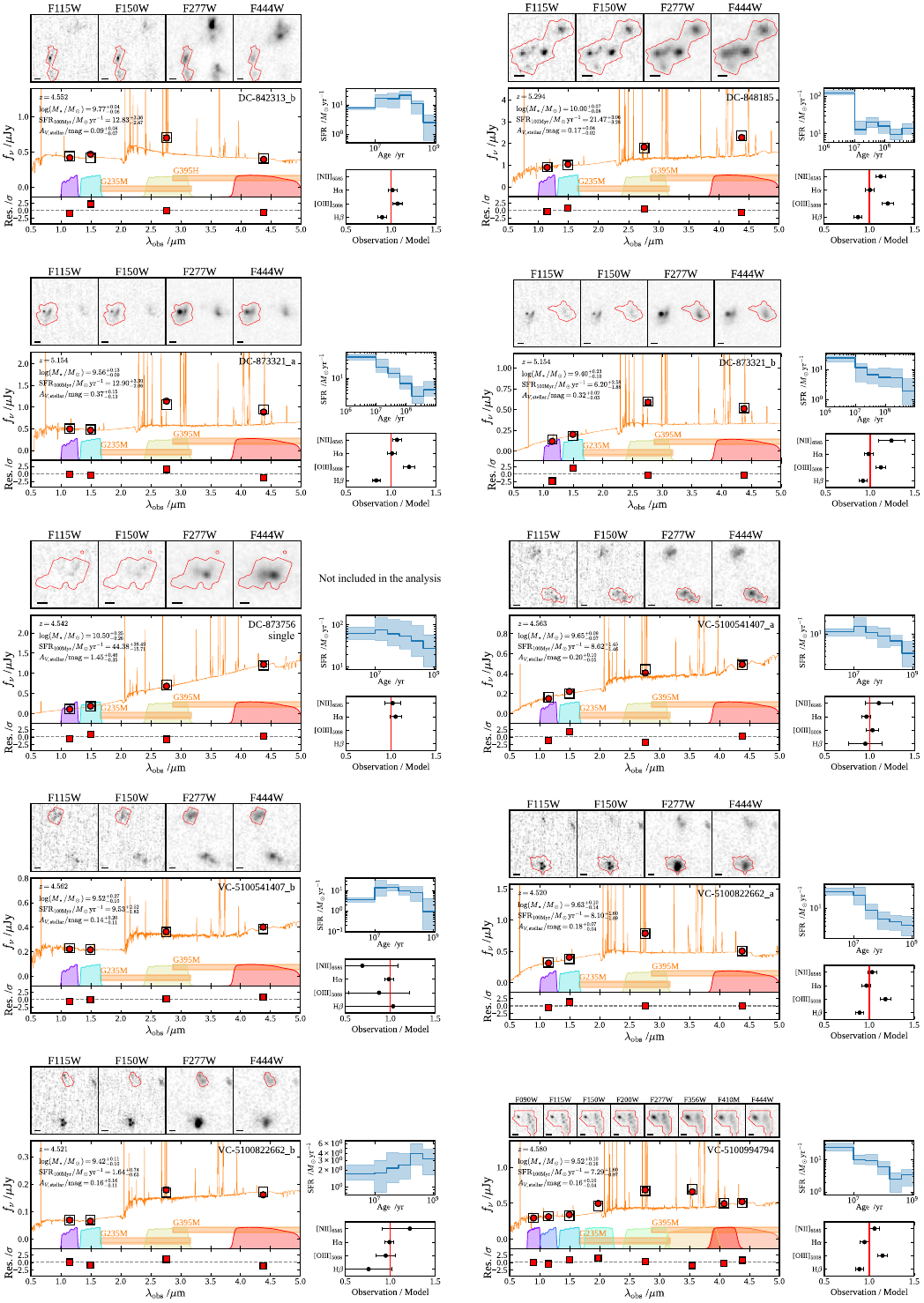}
\caption{(continued)}
\end{figure*}
\begin{figure*}[!h]
\centering
\includegraphics[width=0.9\textwidth]{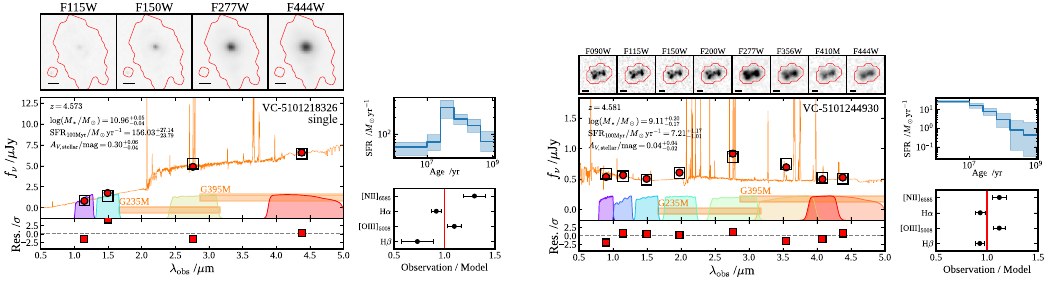}
\caption{(continued)}
\end{figure*}

\section{Integrated SED fits with and without ALMA} \label{sec:appe_sed_ALMA}
In our analysis, ALMA photometry was not included in the fiducial integrated SED fitting. 
The commonly adopted energy-balance approach assumes that stellar and dust emission are co-spatial, i.e., that the UV/optical light absorbed by dust is fully re-emitted in the IR. 
However, this assumption can break down especially in high-redshift galaxies where the UV and FIR emission are often spatially decoupled \citep[e.g.,][]{Kokorev2021ApJ...921...40K}. 
In our case, only a single ALMA band is available, meaning that the fit would rely entirely on the energy-balance assumption and would therefore not be robust. 
For this reason, we excluded ALMA from our fiducial fits. 
Nevertheless, we verified that including ALMA photometry typically yields $f$ values broadly consistent within about 10\%, and thus does not affect our main conclusions. 
A detailed discussion on whether or not to include ALMA data is beyond the scope of this paper and will be presented in a separate paper.

\section{Comparison with previous SED fit results} \label{appe:SEDfit_comparison}
We compare our pixel-by-pixel SED fitting results with the stellar masses reported by \cite{Li2024ApJ...976...70L}, who employed \texttt{MAGPHYS} code \cite{dacunha2015} including ALMA, HST, and JWST broad band photometry. 
We find that our stellar masses are systematically smaller, with a median offset of $\sim$0.4 dex. 
While the comparison is not straightforward owing to multiple differences such as pixel selection, aperture definition, SED model, and the set of input data, one possible contributor is that our fits explicitly include spectroscopic line fluxes in addition to broadband photometry.
This allows us to disentangle line contributions from the underlying stellar continuum, which cannot be achieved with photometry alone. 
Indeed, \cite{Yuan2019A&A...631A.123Y} demonstrated that including line fluxes reduces systematic discrepancies in stellar mass estimates (their average offset is 0.11 dex). 
Nevertheless, we stress that this effect alone is unlikely to account for the full offset, and other factors may also play a role.

\bibliography{reference}{}
\bibliographystyle{aasjournal}

\end{document}